\documentclass[fleqn,usenatbib]{mnras}

\usepackage{newtxtext,newtxmath}

\usepackage[T1]{fontenc}

\defcitealias{K21}{K21}
\defcitealias{D22}{D22}
\defcitealias{L21}{L21}

\def\hscpipe{{\tt{hscpipe v8.4}}}
\def\lens{{\textsc{Lenstronomy}}}

\def\sid{\textit{$\rm S\acute{e}rsic\ index$}}

\def\ss{\textit{$\rm S\acute{e}rsic$}}

\def\fr{\textit{$f_{\rm gal}$}}
\def\msun{\textit{$M_\odot$}}
\def\ergs{\textit{$\rm erg\ s^{-1}$}}
\def\cm{\textit{$\rm cm^{-2}$}}

\def\re{\textit{$R_{\rm e}$}}

\def\fr{\textit{$f_{\rm gal}$}}

\def\lx{\textit{$L_{\rm X}$}}
\def\nh{\textit{$N_{\rm H}$}}
\def\lognh{\textit{${\rm log}\,N_{\rm H}$}}
\def\loglx{\textit{${\rm log}\,L_{\rm X}$}}
\def\loglbol{\textit{${\rm log}\,L_{\rm bol}$}}
\def\logm{\textit{${\rm log}\,\mathcal{M}_\star$}}
\def\mbh{\textit{$\mathcal{M}_{\rm BH}$}}
\def\mbulge{\textit{$\mathcal{M}_{\rm bulge}$}}

\def\logmbh{\textit{${\rm log}\,\mathcal{M}_{\rm BH}$}}

\def\m{\textit{$\mathcal{M}_\star$}}
\def\zspec{\textit{$z_{\rm spec}$}}
\def\zphot{\textit{$z_{\rm phot}$}}
\def\av{\textit{$A_{\rm V}$}}

\DeclareRobustCommand{\VAN}[3]{#2}
\let\VANthebibliography\thebibliography
\def\thebibliography{\DeclareRobustCommand{\VAN}[3]{##3}\VANthebibliography}

\usepackage{graphicx}	
\usepackage{amsmath}	

\title[Host-galaxy of eFEDS AGNs]{The eROSITA Final Equatorial-Depth Survey (eFEDS): Host-galaxy Demographics of X-ray AGNs with Subaru Hyper Suprime-Cam}

\author[Junyao Li et al.]{Junyao Li,$^{1,2}$\thanks{E-mail: junyaoli@illinois.edu}
John D. Silverman,$^{2,3}$
Andrea Merloni,$^{4}$
Mara Salvato,$^{4,5}$
Johannes Buchner,$^{4}$
Andy Goulding,$^{6}$
\newauthor
Teng Liu,$^{4}$
Riccardo Arcodia,$^{4}$
Johan Comparat,$^{4}$
Xuheng Ding,$^{2}$
Kohei Ichikawa,$^{4,7,8}$
Masatoshi Imanishi,$^{9,10}$
\newauthor
Toshihiro Kawaguchi,$^{11}$
Lalitwadee Kawinwanichakij,$^{2}$
and Yoshiki Toba$^{9,12,13,14}$
\\
$^{1}$Department of Astronomy, University of Illinois at Urbana-Champaign, Urbana, IL 61801, USA\\
$^{2}$Kavli Institute for the Physics and Mathematics of the Universe, The University of Tokyo, Kashiwa, Japan 277-8583 (Kavli IPMU, WPI)\\
$^{3}$Department of Astronomy, School of Science, The University of Tokyo, 7-3-1 Hongo, Bunkyo, Tokyo 113-0033, Japan\\
$^{4}$Max-Planck-Institut f\"ur Extraterrestrische Physik (MPE), Giessenbachstrasse 1, 85748 Garching bei M\"unchen, Germany\\
$^{5}$Exzellenzcluster ORIGINS, Boltzmannstr. 2, 85748 Garching, Germany\\
$^{6}$Department of Astrophysical Sciences, Princeton University, 4 Ivy Lane, Princeton, NJ 08544, USA\\
$^{7}$Frontier Research Institute for Interdisciplinary Sciences, Tohoku University, Sendai 980-8578, Japan\\
$^{8}$Astronomical Institute, Graduate School of Science Tohoku University, 6-3 Aramaki, Aoba-ku, Sendai 980-8578, Japan\\
$^{9}$National Astronomical Observatory of Japan, 2-21-1 Osawa, Mitaka, Tokyo 181-8588, Japan\\
$^{10}$Department of Astronomy, School of Science, The Graduate University for Advanced Studies, SOKENDAI, Mitaka, Tokyo 181-8588, Japan\\
$^{11}$Department of Economics, Management and Information Science, Onomichi City University, Onomichi, Hiroshima 722-8506, Japan\\
$^{12}$Department of Astronomy, Kyoto University, Kitashirakawa-Oiwake-cho, Sakyo-ku, Kyoto, Kyoto 606-8502, Japan\\
$^{13}$Academia Sinica Institute of Astronomy and Astrophysics, 11F Astronomy-Mathematics Building, AS/NTU, No.1, Section 4, Roosevelt Road, Taipei 10617, Taiwan\\
$^{14}$Research Center for Space and Cosmic Evolution, Ehime University, 2-5 Bunkyo-cho, Matsuyama, Ehime 790-8577, Japan
}

\date{Accepted XXX. Received YYY; in original form ZZZ}

\pubyear{2015}

\begin{document}
\label{firstpage}
\pagerange{\pageref{firstpage}--\pageref{lastpage}}
\maketitle

\begin{abstract}
We investigate the physical properties, such as star-forming activity, disk vs. bulge nature, galaxy size, and obscuration of 3796 X-ray selected AGNs ($42.0<\loglx/\ergs<44.5$) at $0.2<z<0.8$ in the eFEDS field. Using Subaru Hyper Suprime-Cam imaging data in the $grizy$ bands for SRG/eROSITA-detected AGNs, we measure the structural parameters for AGN host galaxies by performing a 2D AGN-host image decomposition. We then conduct spectral energy distribution fitting based on the decomposed galaxy emission to derive stellar mass ($9.5<\logm/\msun<12.0$) and rest-frame colors for AGN hosts. We find that (1) AGNs can contribute significantly to the total optical light down to $\loglx\sim42.5\ \ergs$, thus ignoring the AGN component can significantly bias the structural measurements; (2) AGN hosts are predominately star-forming galaxies at $\logm \lesssim 11.3\,\msun$;
   (3) the bulk of AGNs (64\%) reside in galaxies with significant stellar disks (\ss~index $n<2$), while their host galaxies become increasingly bulge dominated ($n\sim4$) and quiescent at $\logm \gtrsim 11.0\,\msun$; (4) the size--stellar mass relation of AGN hosts tends to lie between that of inactive star-forming and quiescent galaxies, suggesting that the physical mechanism responsible for building the central stellar density also efficiently fuel the black hole growth; (5) the hosts of X-ray unobscured AGNs are biased towards face-on systems and the average $E(B-V)/\nh$ is similar to the galactic dust-to-gas ratio, suggesting that some of the obscuration of the nuclei could come from galaxy-scale gas and dust, which may partly account for (up to 30\%) the deficiency of star-forming disks as host galaxies for the most massive AGNs. These results are consistent with a scenario in which the black hole and galaxy grow in mass while transform in structure
and star-forming activity, as desired to establish the local scaling relations. Our results highlight the importance of 2D image decomposition for optical studies of AGNs and their host galaxies, and strengthen equivalent findings for the hosts of optically-selected SDSS quasars as recently reported based on HSC imaging.
\end{abstract}


\section{Introduction}
X-ray surveys of the extragalactic sky, primarily with {\it{Chandra}} and {\it{XMM-Newton}}, have made remarkable progress in the study of accretion onto supermassive black holes (SMBHs) across our observable universe \citep[see][for a review]{BrandtHasinger2005}. We now have a clearer understanding of the cosmic evolution of Active Galactic Nuclei (AGNs) from $z\sim6$ to the present with the inclusion of the obscured population \citep[e.g.,][]{Merloni2014}. Remarkably, the average accretion rate density of SMBHs mirrors that of the cosmic star formation rate (SFR) density lending evidence for a connection between the growth of SMBHs and their host galaxies \citep[e.g.,][]{Silverman2008, Madau2014, Aird2015, Yang2018}. With an assumption on the accretion efficiency, the integral of the mass growth rate of SMBHs with cosmic time broadly agrees with the local mass density of inactive SMBHs \citep{Soltan1982, Shankar2013}.

Over the past decade or more, large observational efforts from both the ground and space have accelerated investigations into the properties of the host galaxies of distant X-ray detected AGNs motivated by the possible links mentioned above, the local SMBH-bulge scaling relations \citep{KormendyHo2013}, and the need for energy injection from SMBHs (i.e., AGN feedback) to produce realistic massive galaxies in cosmological simulations \citep[e.g.,][]{Silk1998, Sijacki2007, Weinberger2018}. It is well known that X-ray selection of AGNs results in cases that are underluminous in the optical thus facilitating the views of their underlying host galaxy. However, challenges with disentangling the AGN and galaxy emission are always an issue at most wavelengths with the likely exception of the FIR. Based on these efforts, we firmly understand that the bulk of X-ray AGNs prefer to reside in massive star-forming galaxies \citep[e.g.,][]{Silverman2009, Brusa2009, Suh2017, Li2019, Zou2019, Florez2020} and within dark matter halos near the peak efficiency in converting baryons to stars \citep[e.g.,][]{Gilli2009, Allevato2011, Georgakakis2019}. Their hosts show signs of having a light profile descriptive of disk galaxies \citep[e.g.,][]{Schawinski2011, Cisternas2011a} and more compact emission \citep[e.g.,][]{Kocevski2017, Silverman2019, Ni2021, L21} that may indicate the emergence of a central stellar concentration (i.e., bulge). At $z < 1$, galaxy mergers can effectively trigger AGN accretion but do not appear to be the dominant mechanism (e.g., \citealt{Cisternas2011a, Silverman2011, Kocevski2012, Goulding2018, Ellison2019}; Tang et al. submitted), while in the gas-rich universe at higher redshifts, the picture is less clear \citep[e.g.,][]{Mechtley2016, Shah2020, Silva2021}. Even with such progress, we have not yet demonstrated a causal link between the galaxies, their evolution, and the activation and feedback of their central SMBHs.

Some limitations may be due to the fact that current X-ray selected AGN samples, sufficient in depth to probe the higher redshift universe and faint obscured population \citep[e.g.,][]{Hasinger2007, Elvis2009, Luo2017}, do not reach the level of statistical significance comparable to wide-field studies of AGNs in the low-redshift universe  as accomplished with the Sloan Digital Sky Survey \citep[e.g.,][]{Kauffmann2003,Heckman2006,Lyke2020}. We now have a better understanding that large number statistics ($>10^3$ per redshift interval $dz\sim0.1$) are required to break degeneracies between parameters (e.g., redshift, stellar mass, environment, galaxy structure, galaxy color) inherent in studies of the connection between SMBHs and galaxies. Wide-area X-ray imaging is needed to fully exploit the rich optical and near-infrared imaging currently being amassed by Subaru Hyper Suprime-Cam \citep[][]{Aihara2022}, the Dark Energy Survey and future efforts with the Vera Rubin Observatory (LSST), Euclid, and the Nancy Grace Roman Space Telescope (NGRST). With the successful launch of eROSITA \citep{Merloni2020, Predehl2021} which uniquely combines a higher sensitivity than its predecessor ROSAT survey, a large field of view ($\approx 1$ degree in diameter)  and a large collecting area designed for deep all-sky survey, large samples of X-ray detected AGNs across cosmic time are now being amassed.

Here, we use the wide and deep optical imaging of the HSC Subaru Strategic Program to report on the properties of the host galaxies of X-ray AGNs detected by eROSITA. The joint effort between the eROSITA and Subaru/HSC teams focuses on a 140~deg$^2$ region of the sky observed during the SRG/eROSITA Performance Verification (PV) phase, i.e., the eROSITA Final Equatorial Depth Survey (eFEDS), where the deep five-band ($grizy$) imaging from HSC can be exploited for a range of science from eROSITA. Here, we implement 2D image analysis tools to decompose the AGN and host galaxy optical emissions (while considering the point-spread function; PSF) for 3796 AGNs at $0.2<z<0.8$ to study the properties of their host galaxies including stellar mass, disk vs. bulge nature, rest-frame color (indicative of the SFR or age of the underlying population), and obscuration. We attempt to place X-ray AGNs in the broader context of galaxy evolution by comparing to a large sample of inactive (comparison) galaxies available from the HSC-SSP with the equivalent intrinsic measurements (\citealt{K21}; hereafter \citetalias{K21}). Furthermore, we compare to a recent study on the hosts of optically-selected quasars from SDSS using the HSC imaging and tools used in this study (\citealt{L21}; hereafter \citetalias{L21}). This paper is structured as follows. Section \ref{sec:sample} describes the X-ray AGN selection and their optical counterparts.  Section \ref{sec:method} introduces our image decomposition method. The physical properties of AGNs and their host galaxies are presented in Section \ref{sec:result}. The implications of our results in the broader context of AGN-galaxy connection are discussed in Section \ref{sec:discussion}. In Section \ref{sec:summary} we summarize our main findings. Throughout, we assume $\Omega_\Lambda$ = 0.7, $\Omega_{\rm m}$ = 0.3, and $H_0 = 70\rm\ km\ s^{-1}\ Mpc^{-1}$. Magnitudes are given in the AB system.

\section{X-ray AGN selection and optical counterparts}
\label{sec:sample}

The eFEDS point source catalog \citep{Brunner2022} enables us to construct a large sample of AGNs for studies of their host galaxies. In total, the catalog contains 27910 primary X-ray point sources with a detection likelihood (i.e., signal-to-noise) greater than 6 in the 0.2--2.3 keV band. The spurious detection rate has been assessed at the level of 1.8\%. The X-ray flux limit of the eFEDS survey is $\sim6.5\times10^{-15}\ \rm erg\ s^{-1}\ cm^{-2}$ in the soft (0.5--2~keV) band, for a completeness level of $\sim80\%$.

Optical-to-mid-IR counterparts of X-ray point sources are identified in \cite{Salvato2022} through the Bayesian-based {\tt{NWAY}} method and the maximum likelihood ratio-based {\tt{astromatch}} method using the data taken from the DESI Legacy Imaging Surveys Data Release 8 \citep{Dey2019}. A high fraction ($90.5\%$) of X-ray sources have reliable optical associations, with each object being assigned a counterpart reliability flag ({\tt{CTP\_quality}}). The contamination is primarily due to uncertainties on the centroid of the X-ray emission which is typically around $3^{\prime\prime}-5^{\prime\prime}$ dependent on the X-ray source significance and off-axis angle \citep{Brunner2022}. About 80\% of the X-ray point sources have {\tt{CTP\_quality} $\geq 2$}, meaning that the counterpart has been identified as the best association by at least one method, and are classified as {\tt{"Likely"}} or {\tt{"Secure"}} extragalactic by assessing their X-ray, optical, and IR properties. 
Spectroscopic redshifts (\zspec) are available for 6591 sources from past spectroscopic surveys (mainly from SDSS; \citealt{Ahumada2020, Abdurrouf2022}). Photometric redshifts (\zphot) for eFEDS sources are available in \cite{Salvato2022}, which are measured from multi-wavelength SED fitting using {\tt{LePHARE}}. 

\cite{Liu2022} performed X-ray spectral fitting for eFEDS sources to characterize their properties. The X-ray spectral fitting is performed with {\tt{BXA}} \citep{Buchner2014}, which connects {\tt{XSPEC}} \citep{Arnaud1996} with the {\tt{UltraNest}} nested sampling algorithm \citep{Buchner2017}.  The fitting starts from a single absorbed powerlaw model ({\tt{TBabs*zTBabs*powerlaw}}), as most of the X-ray sources are expected to be AGNs. A second soft powerlaw or a blackbody component is added for sources with at least 20 net counts to account for the possible existence of a soft-excess component. As some of the X-ray sources might be stars or compact galaxy clusters, the collisionally-ionized gas emission model {\tt{apec}} is also performed. The outcome of this X-ray spectral analysis is an eFEDS AGN catalog  which comprises 22079 eFEDS point sources, as well as the intrinsic AGN properties such as absorption corrected rest-frame 0.5--2 keV luminosity  ({\tt{LumiIntr\_Med\_s}}; \lx) and the hydrogen column density ({\tt{lognH\_s\_m1}}; \nh) from a single-powerlaw fit. 

We build our AGN sample from the \cite{Liu2022} catalog at $0.2<z<0.8$ since the AGN-host decomposition with HSC imaging is most reliable in this redshift range \citepalias{L21}. The selected sources automatically satisfy {\tt{CTP\_quality} $\geq 2$} and {\tt{"Likely"}} or {\tt{"Secure"}} extragalactic as required in \cite{Liu2022}. We  exclude 691 sources that lie outside the inner 90\% area of the eFEDS footprint ({\tt{inArea90 = True}}) as the data quality significantly drops due to higher background, stronger vignetting, and shorter exposure. To ensure reliable \zphot, we only include sources with a level-4 reliability, which means the \zphot~in \cite{Salvato2022} is consistent with those derived from the deep neural network technique of Nishizawa et al. (in prep) for sources in common. This leads to 4975 X-ray AGNs as our parent sample. We then cross-match their optical counterparts with the HSC S20A wide layer data using a 1\arcsec~matching radius to identify eROSITA AGNs that have clean HSC images in the five optical bands based on the following criteria: 

\begin{enumerate}
    \item isprimary = True
    \item nchild = 0 
    \item pixelflags\_edge = False
    \item pixelflags\_bad = False
    \item pixelflags\_crcenter = False
    \item mask\_brightstar\_halo, ghost, blooming = False
    \item pixelflags\_saturatedcenter = False
\end{enumerate}

\noindent These selection criteria ensure the following for the HSC-detected optical counterparts to eFEDS sources: identification of a unique source after deblending (1--2), away from the edge of a CCD (3), contains no bad pixels or defects from cosmic rays at their cores (4--5), away from nearby bright stars (6), and unsaturated in the image center (7). 

\begin{figure}
\includegraphics[width=\linewidth]{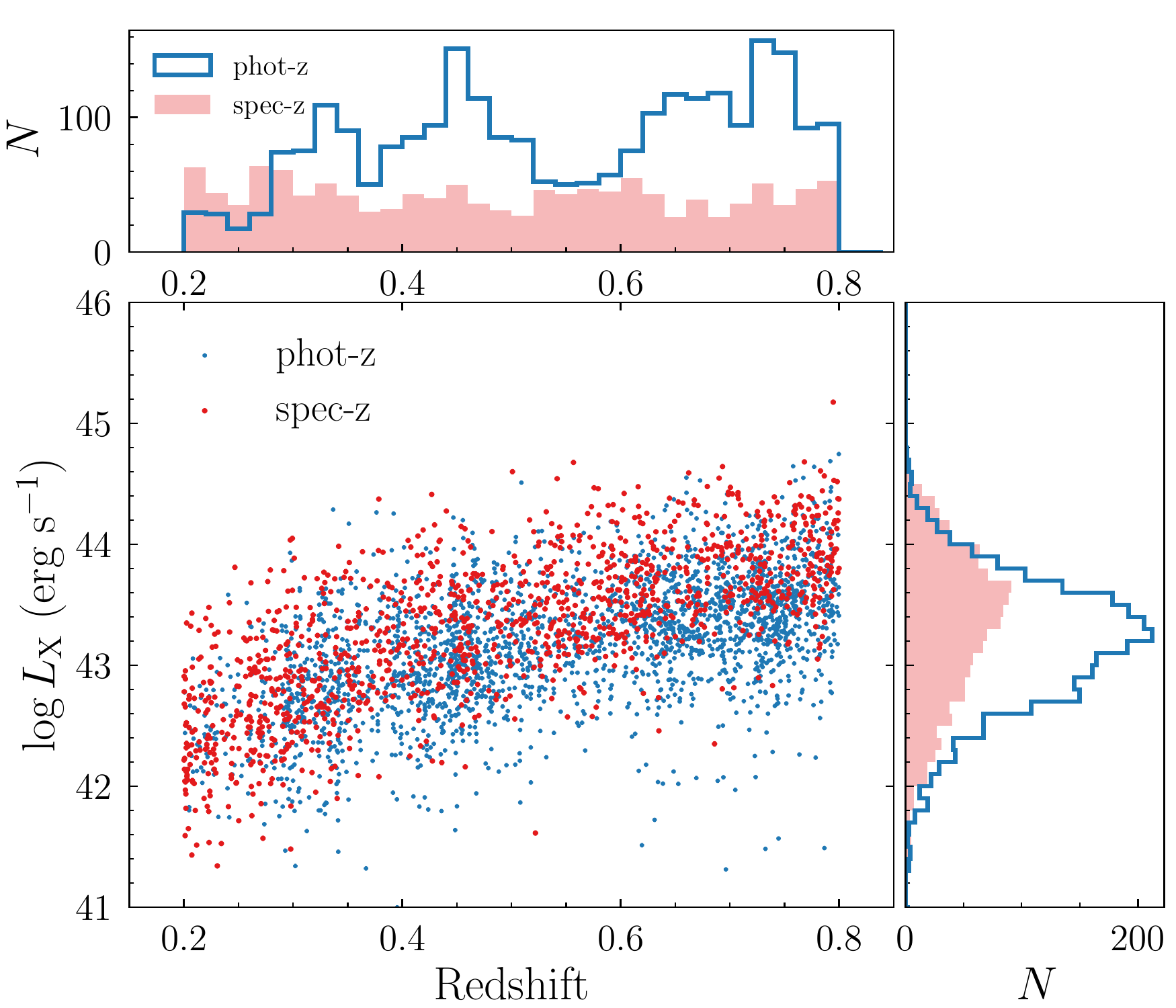}
\caption{Absorption-corrected rest-frame 0.5--2.0~keV X-ray luminosity as a function of redshift for eFEDS AGNs at $0.2<z<0.8$. Those having spectroscopic and photometric redshifts are indicated.}
\label{fig:z}
\end{figure}

The redshift and \lx~distributions for 3796 matched sources are shown in Figure \ref{fig:z}. Over 95\% of our sample have \lx~larger than $10^{42}\ \rm erg\ s^{-1}$, thus are expected to be bona-fide X-ray AGNs. Adopting the bolometric correction in the 0.5--2~keV band given in \cite{Lusso2012}, the bolometric luminosity of our sample spans from $10^{43}\,\ergs$ to $10^{46}\,\ergs$ with a mean (median) of $8\times10^{44}\,\ergs$ ($3\times10^{44}\,\ergs$). Most (3509/3796) of them have $\nh < 10^{22}\,\cm$ due to that eROSITA is most sensitive to the soft X-ray emissions.

\begin{figure*}
\centering
\includegraphics[width=0.7\linewidth]{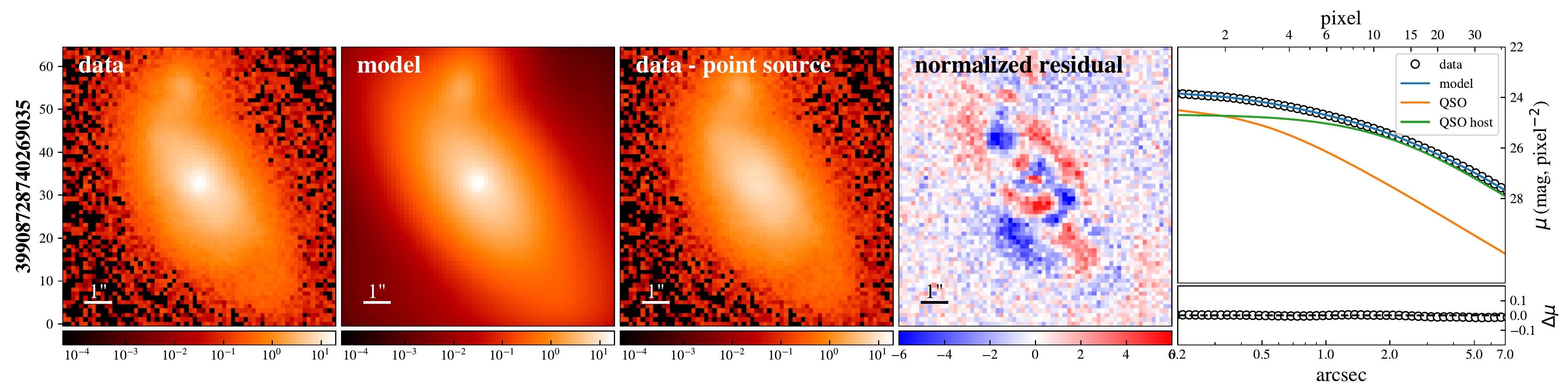}

\includegraphics[width=0.7\linewidth]{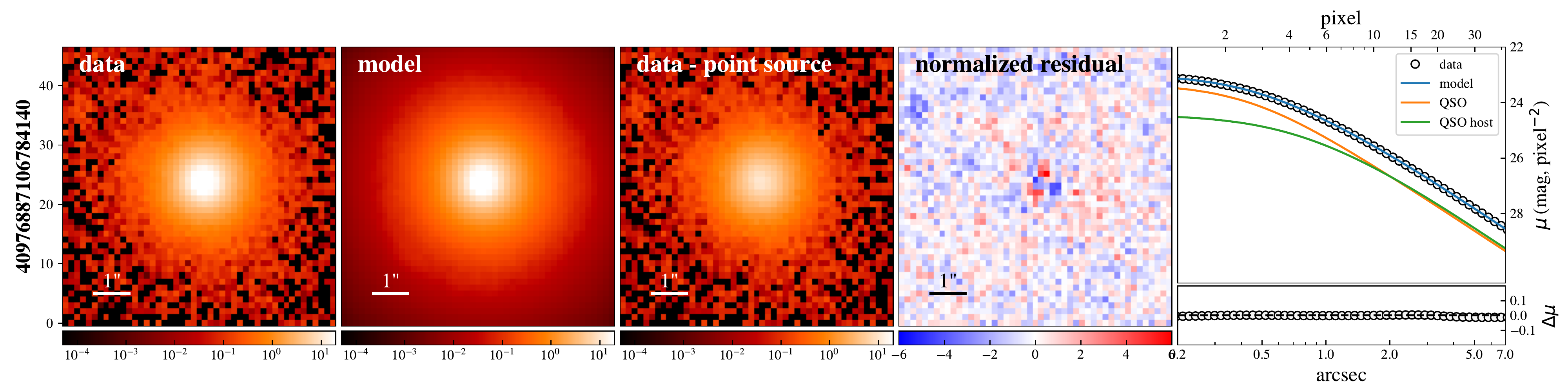} 

\includegraphics[width=0.7\linewidth]{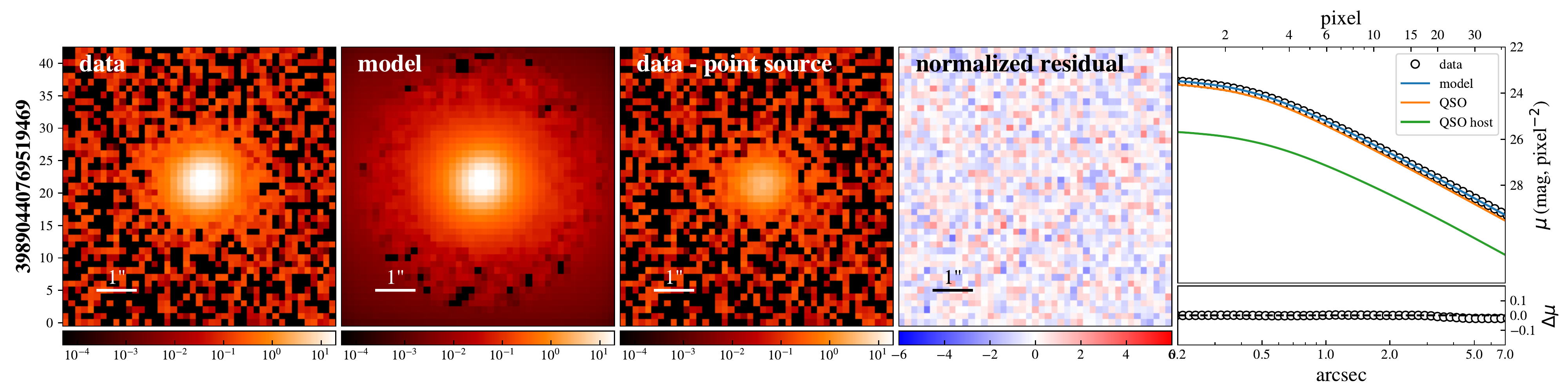}
\caption{2D image decomposition of representative eFEDS AGNs based on HSC $i$-band imaging. The panels from top to bottom are: (1) a $z=0.27$ AGN with $\fr=89\%$; (2) a $z=0.53$ AGN with $\fr=41\%$; (3) a $z=0.63$ AGN with $\fr=25\%$. The panels from left to right are as follows: (1) observed HSC $i$-band image; (2) best-fit point source + host-galaxy + nearby companions (if present) model convolved with the PSF; (3) data minus the point source model (the pure-galaxy image); (4) fitting residual divided by the error map; (5) 1D surface brightness profiles (top) and the corresponding residual (bottom).}
\label{fig:example}
\end{figure*}

\section{Method: 2D optical image decomposition}
\label{sec:method}

We conduct a 2D decomposition of the total optical emission into the AGN and host galaxy contribution separately using the HSC imaging data in five bands ($grizy$) as fully presented in \citetalias{L21}, which has been successfully implemented to measure the host galaxy properties of $\sim 5000$ SDSS quasars. We refer the readers to Sections 3 and 4 in \citetalias{L21} for the details of our decomposition method and reliability checks through extensive image simulations (based on $\sim1$ million simulated model galaxies and $\sim5000$ real CANDELS galaxies plus model AGNs with realistic noise added) which include dependencies on host-to-total emission, magnitude, galaxy structural properties, and resolution (seeing conditions). Here, we briefly summarize the concept of our method.

We generate background-subtracted co-added HSC images ($60 \times 60^{\prime\prime}$) in $grizy$ bands \citep{Kawanomoto2018}, together with the variance images and PSF models for each source using the HSC pipeline \hscpipe~ \citep{Bosch2018}. The size of the cut-out image used for decomposition is chosen automatically to be between $41×\times41$ ($7^{\prime\prime} \times 7^{\prime\prime} $) and $131 \times 131$ ($22^{\prime\prime} \times 22^{\prime\prime} $) pixels to ensure that the flux from the target AGN and close companions are fully included.
We then implement an automated fitting routine using the image modeling tool \lens~\citep{Birrer2015, Birrer2018} to decompose the cut-out images for the target AGN into a point source and an underlying host galaxy, assuming that the point source can be described by a PSF model. A 2D \ss~profile  convolved with the PSF is used to model the host-galaxy component. If present, companion objects, brighter than 25~mag in each cutout image, are simultaneously modeled with a \ss~profile. We first fit each source using the $i$-band image since it was observed under the best seeing conditions (typically 0.6\arcsec)  and that the host-galaxy contribution in the $i$-band is relatively high. Then sources are fit in the other bands by fixing the structural parameters to the $i$-band results. The outputs of our fitting routine include the model AGN and galaxy magnitude in all optical bands, and the structural properties of the host galaxy in the $i$-band, i.e., \ss~index ($0.3<n<7.0$), galaxy size (half-light radius of the semi-major axis $0.1^{\prime\prime}<\re<5.0^{\prime\prime}$), ellipticity ($0<\epsilon<0.8$) and position angle. Motivated by \citetalias{L21}, we further fix the \ss~index to $n=0.7$ and $n=2.0$ for those hit the lower and upper boundaries, respectively, as such objects are most likely to have moderate \ss~indices but the model degeneracy or/and the PSF mismatch led to very low/high \ss~indices.
In Figure~\ref{fig:example} we show results from the image decomposition for three representative AGNs with varying host-to-total flux ratio (\fr) in the $i$-band. 

\begin{figure*}
\centering
\includegraphics[width=\linewidth]{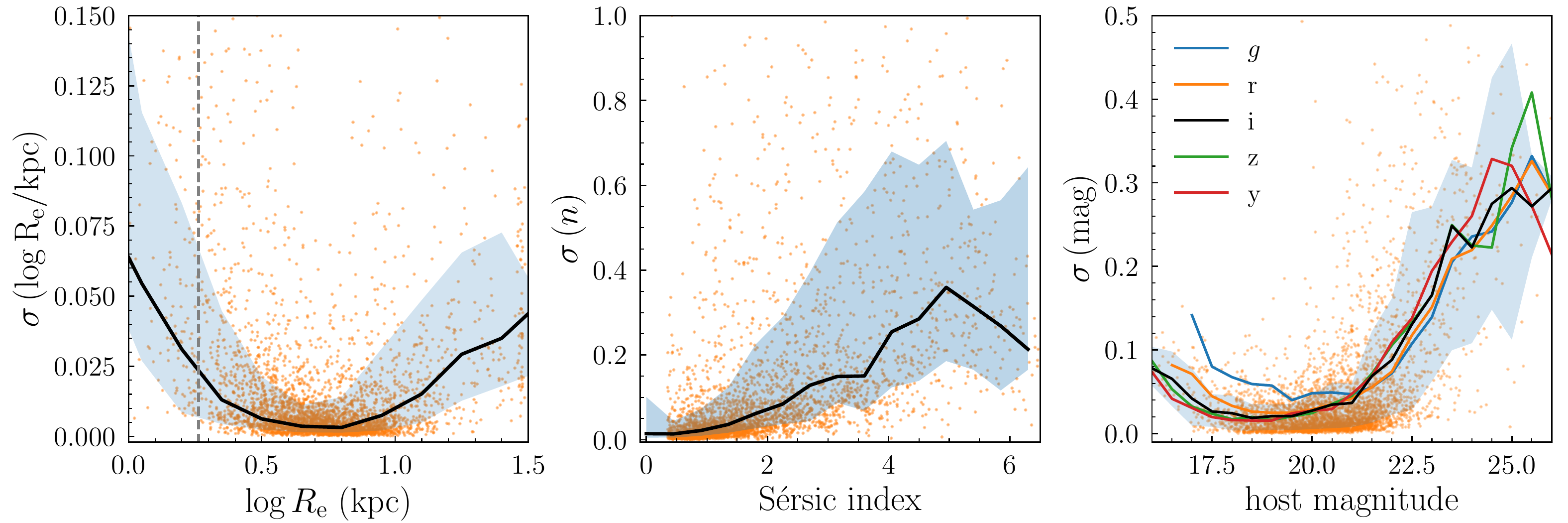}
\caption{Estimated uncertainties on the galaxy size, \ss~index and the decomposed host-galaxy magnitudes. The blue shaded region (only show for the $i$-band in the right panel) and colored curves represent the 16-50-84th percentiles of the error distribution. The vertical dashed line in the left panel indicates where the size measurements are below the $i$-band resolution limit ($R_{\rm e} < 0.5\rm \times FWHM$) at $z=0.5$.}
\label{fig:error}
\end{figure*}

We then adopt the image simulation result in \citetalias{L21} to correct for systematic measurement biases on the decomposed galaxy magnitude which arises from the fact that the structure parameters are wavelength dependent but are fixed to the $i$-band values. 
The uncertainties of our decomposition results are evaluated by examining the scatter between the input and output parameters in the image simulations at a given host galaxy magnitude and host-to-total flux ratio for each source. In Figure \ref{fig:error} we show the estimated uncertainties for the three main parameters: galaxy size, \ss~index, and decomposed galaxy magnitudes. The uncertainties on the galaxy size are significantly higher for the most compact galaxies whose sizes are smaller than the HSC PSF, and the most extended galaxies whose outskirts are indistinguishable from the background light. The uncertainties on the \ss~index increase when galaxies moving from disk-dominated to bulge-dominated due to the difficulty of separating the point source from a centrally concentrated bulge. The host galaxy magnitudes are well constrained for bright objects ($i \lesssim 23.0$), while the faint objects only contribute to a small fraction of our sample. Note that these uncertainties are estimated under a certain model, thus they could be underestimated due to model mismatch (e.g., PSF mismatch, existence of complex galaxy structures).

Spectral energy distribution (SED) fitting is then performed on the decomposed galaxy flux (corrected for measurement bias and galactic extinction) to infer the stellar population properties for AGN host galaxies such as stellar mass and rest-frame colors using CIGALE \citep{Boquien2019, Yang2020}.  A \cite{Chabrier2003} initial mass function, a \cite{Bruzual2003} stellar population model, a delayed star-formation history and a \cite{Calzetti2000} extinction law are assumed. In the fitting, we add an additional 10\% flux error in quadrature to account for the unknown uncertainties in the decomposition due to model mismatch. This results in a typical uncertainty on $\logm$ of 0.25 dex. We refer the readers to Sections 3.2 and 4.2 in \citetalias{L21} for a full description of our method and the evaluation of the robustness of the SED fitting results through simulations. 

Following \citetalias{L21}, we set a stellar-mass cut ($\mathcal{M}_{\star,\rm cut}$) to construct a relatively mass-complete sample, which is chosen to correspond to the 90\% completeness limit at an $i$-band magnitude of 23~mag using the method described in \cite{Pozzetti2010}.  This is to avoid problematic decomposition and stellar mass measurements for faint host galaxies. Mass limits at $z=0.3$, 0.5, and 0.7 are $10^{9.2}\,M_\odot$, $10^{9.8}\,M_\odot$, and  $10^{10.3}\,M_\odot$, respectively. This leaves 3414 sources for the following analysis.

\section{Results}
\label{sec:result}

\subsection{Decomposed AGN and Galaxy Emission}
\label{subsec:ratio}

\begin{figure}
\includegraphics[width=\linewidth]{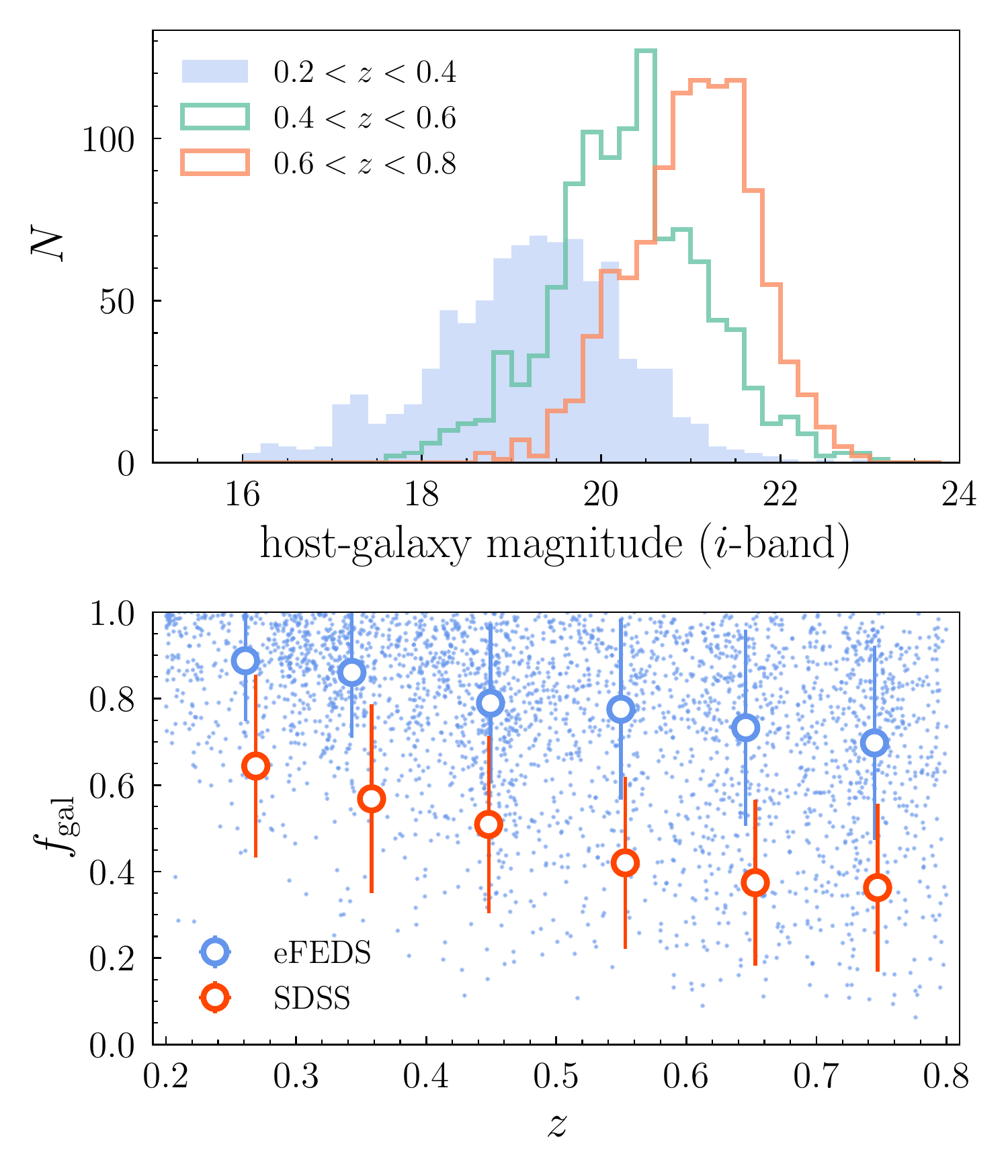}
\caption{Top: decomposed host-galaxy magnitude in three redshift intervals measured in the HSC $i$-band. Bottom: fraction of the galaxy light (\fr) to the total AGN + galaxy emission measured in the $i$-band as a function of redshift. Mean values and their dispersions are shown by open circles and error bars. For comparison, we plot the equivalent values for optically-selected SDSS type 1 quasars from \citetalias{L21}. This illustrates that X-ray selected AGNs have a greater host galaxy contribution to the total optical emission over all redshifts considered.}
\label{fig:ratio}
\end{figure}

\begin{figure*}
\includegraphics[width=\linewidth]{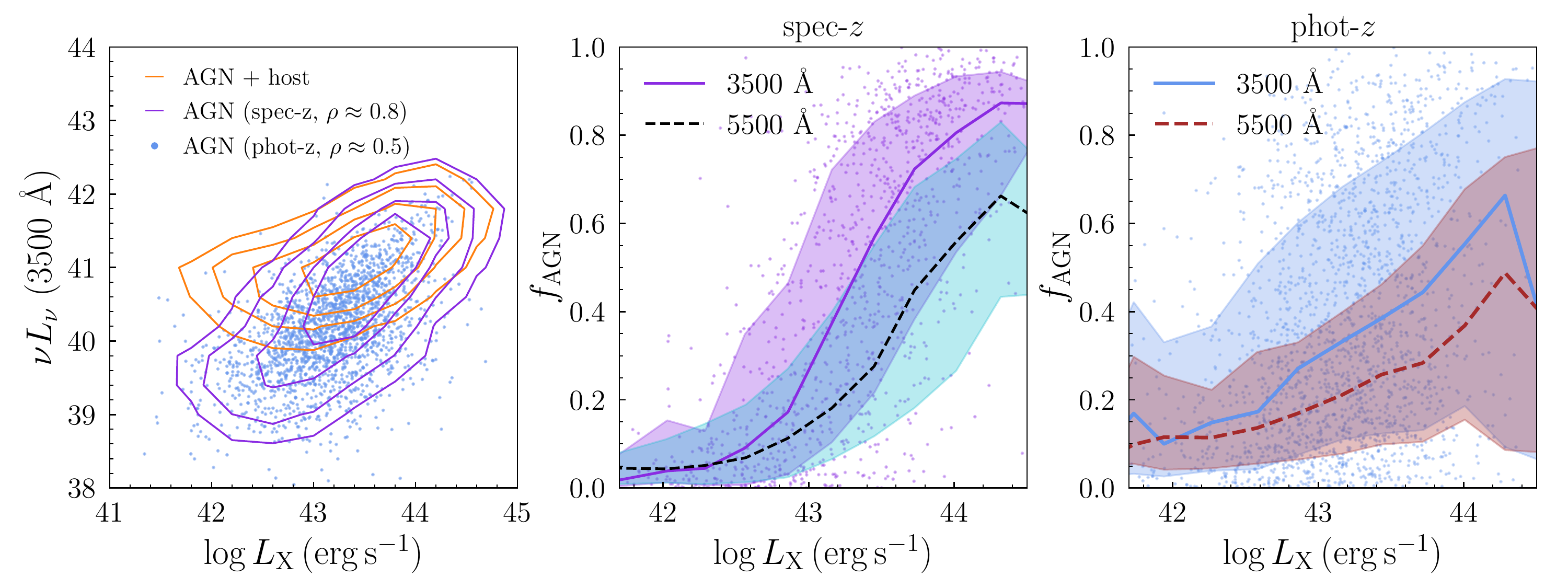}
\caption{Left: rest-frame optical luminosity at 3500~\AA~as a function of \hbox{X-ray} luminosity. Orange contours show the total optical luminosity prior to decomposition. Clean (i.e., host free) AGN luminosities are shown by purple contours (spec-$z$ sample) and blue points (photo-$z$ sample). Middle and Right: fractional contribution of AGN light to the total emission ($f_{\rm AGN}$) at rest-frame 3500~\AA~and 5500~\AA~as a function of X-ray luminosity. Median values and their 16th -- 84th dispersions are shown by colored curves and shaded regions, respectively.}
\label{fig:optical_xray}
\end{figure*}

The decomposed host-galaxy magnitude and the fraction of the host-galaxy light to the total emission (\fr) measured in the HSC $i$-band as a function of redshift is shown in Figure \ref{fig:ratio}. X-ray-selected eFEDS AGNs have a stronger galaxy contribution to the optical light compared to SDSS quasars in \citetalias{L21} due to lower AGN luminosities and higher extinctions of the nuclei (see Section \ref{subsec:nh}). The average \fr~is about $70 - 90\%$ that decreases with redshift due to the increasing optical luminosity of AGNs. Having such a strong host contribution, we can perform robust measurements of the host-galaxy properties since the uncertainties on the decomposition decrease with increasing \fr~\citepalias{L21}.

To aid in optical studies of AGNs and their host galaxies, we demonstrate the impact of the removal of the host galaxy emission in Figure \ref{fig:optical_xray}. In the left panel, we compare the optical luminosity of the AGN, derived from 2D decomposition, to the X-ray luminosity. We use the $g$-band luminosity at $z<0.6$ and $i$-band luminosity at $z>0.6$ to probe the emission at rest-frame $\sim3500$~\AA. The galaxy+AGN luminosity for the full sample, prior to decomposition, is shown as orange contours. The decomposed AGN luminosity for the spec-$z$ and photo-$z$ samples are shown by purple contours and blue points, respectively. 
After correcting for the host contamination, a clear positive correlation between optical and X-ray luminosities is seen with Spearman's correlation coefficients $\rho\approx0.8$ and $\rho\approx0.5$ for the spec-$z$ and photo-$z$ samples, respectively, thus validating our decomposition result. 

In the middle (spec-$z$ sample) and right (photo-$z$ sample) panels we show the individual and average AGN contribution to the optical light at rest-frame 3500~\AA~and 5500~\AA~(probed by $i$-band at $z<0.6$ and $z$-band at $z>0.6$) as a function of X-ray luminosity. As expected, the AGN fraction decreases with \lx. However, we notice that the $L_{\rm 3500}\,(\fr)-\lx$ relation for the photo-$z$ sample is flatter than the spec-$z$ sample. This is likely due to photo-$z$ errors where objects with different redshifts contaminate our measurements. We also find a slight difference in the structural parameters between the two samples, but it is not significant enough to affect our main conclusions. Therefore, we do not further distinguish the two samples in the following analysis.

\begin{figure*}
\centering
\includegraphics[width=0.75\linewidth]{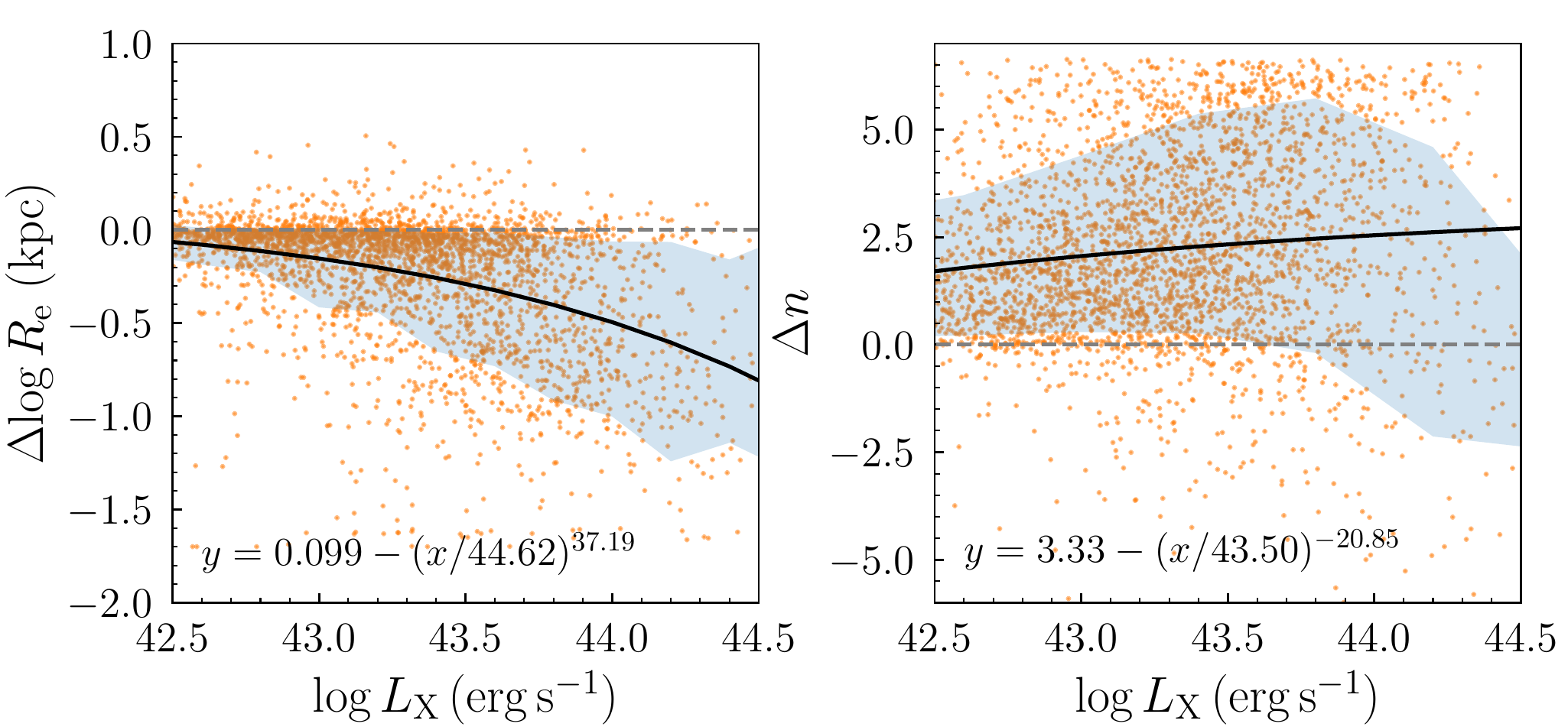}
\caption{Comparison of the difference in the best-fit galaxy size ($\Delta R_{\rm e}$) and \ss~index ($\Delta n$) without and with a point source in the fitting as a function of X-ray luminosity. The average offsets (black curves) are fitted by $y = b - (x/a)^n$ with the best-fit parameters given in the plot. The 16th – 84th dispersions are shown by shaded regions. }
\label{fig:without_ps}
\end{figure*}

Note that although \fr~is positively correlated with \lx, its distribution is broad at a given \lx~and can be as high as $\sim40\%$ down to $\loglx\sim42.5\,\ergs$. Therefore, neglecting the point source component when measuring the host properties may significantly bias the derived structural parameters. To illustrate this, we fit the $i$-band images without adding a point source and compare the derived galaxy size and \ss~index with our default results in Figure \ref{fig:without_ps}. The average offset can be described by a functional form $y = b - (x/a)^n$ and the best-fit parameters are summarized in Figure \ref{fig:without_ps}.
Note that we consistently set $\re > 0.1^{\prime\prime}$ in the fitting although galaxies will be more compact without isolating the point source. It can be clearly seen that the structural measurements are systematically biased even at the lowest X-ray luminosities, especially the \ss~index. Although the size measurements at $\loglx<43\,\ergs$ are less affected, we emphasize that the size difference between AGNs and control galaxies are at the same level (see Section \ref{subsec:size}), thus ignoring these systematics can undermine the comparison between AGNs and inactive galaxies.
Therefore, there is no secure cut on X-ray luminosity where we can be confident to ignore the AGN emission when analyzing their host-galaxy properties in extragalactic surveys, at least for the unobscured population. To the first order, one can use our best-fit offset vs. \loglx~relation to perform corrections on the structural parameters of unobscured AGNs if a decomposition is not performed.

\subsection{Host galaxy: stellar masses and colors}
\label{subsec:color}

\begin{figure*}
\centering
\includegraphics[width=\linewidth]{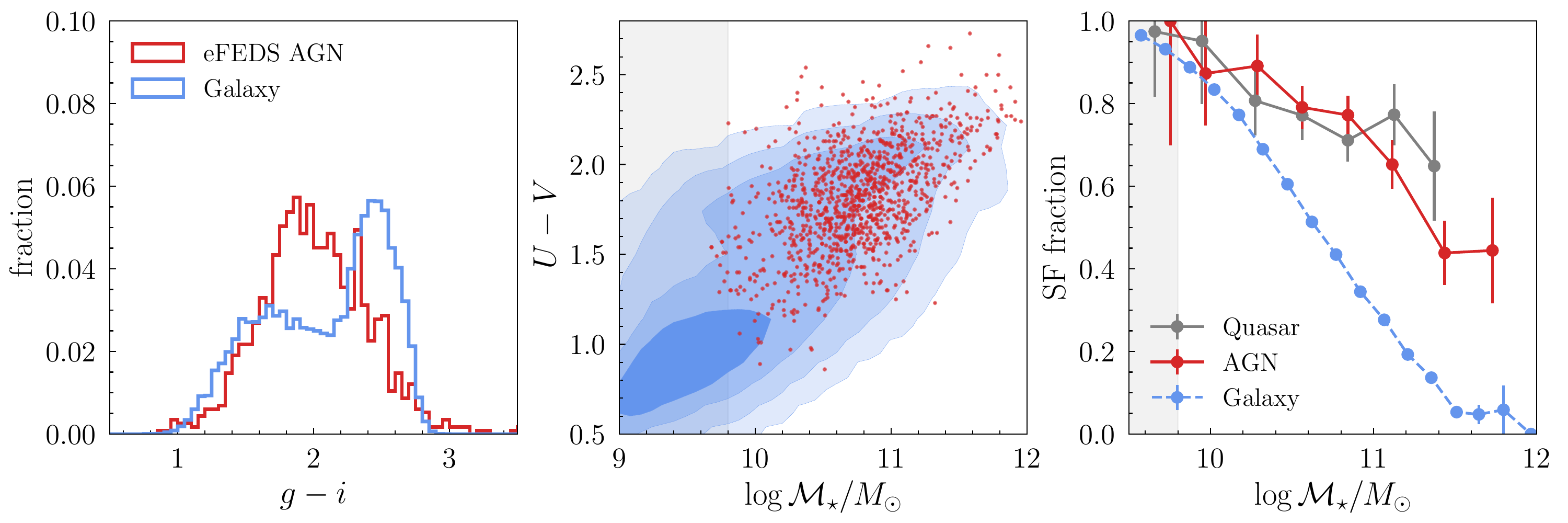}

\caption{Left: distribution of the observed $g-i$ color of AGN host galaxies (red) and stellar mass-matched inactive galaxies (blue).
Middle: rest-frame $U-V$ color of the host galaxy as a function of stellar mass. The eFEDS AGNs and inactive galaxies are shown in red and blue, respectively. 
Right: star-forming fraction as a function of stellar mass for eFEDS AGNs, SDSS quasars, and inactive galaxies. All the three panels are plotted for $0.4<z<0.6$. The vertical shaded region marks our \m~cut at $z=0.5$.}
\label{fig:uvm}
\end{figure*}

In Figure \ref{fig:uvm}, we examine the color distribution of our sample to probe the level of star-forming activity in AGN host galaxies by comparing to a large sample of inactive HSC galaxies in \citetalias{K21}. \citetalias{K21} presented structural analysis for $\sim 1$ million inactive HSC galaxies with structural parameters derived by 2D \ss~profile fitting of HSC $i$-band images. Their photometric redshifts, stellar masses, and rest-frame colors are derived using MIZUKI \citep{Tanaka2015} and EAZY \citep{Brammer2008} with HSC photometry, which assumed the same SED fitting models as our work, thus are suitable for our comparison analysis.

We use the $0.4<z<0.6$ sample as representative while the results are consistent across the whole redshift range. We first compare the observed $g-i$ color (AGN subtracted) with inactive galaxies where the two filters are chosen to probe rest-frame wavelengths below and above the 4000~\AA~break, respectively. As shown in the left panel in Figure \ref{fig:uvm}, the observed color of eFEDS AGNs clearly shifts toward blue galaxies and tend to lie between the two peaks of inactive galaxies. This is confirmed by examining the rest-frame $U-V$ vs. \m~diagram from SED fitting as shown in the middle panel. While AGN hosts are wide spread over the color-\m~plane, the number density of AGNs peaks in a region that is bluer relative to the red sequence, and more massive and redder than the blue cloud. This location can be interpreted as a region of transition from star-forming to quiescent activity. However, there are inherent biases in flux limited AGN samples and the rest-frame colors are dependent on assumptions \citep{Silverman2009}.

To aid in our assessment of AGN host colors relative to the inactive population, we further compare the star-forming fraction among eFEDS AGNs, SDSS quasars in \citetalias{L21}, and inactive HSC galaxies as a function of stellar mass. The classification of star-forming and quiescent galaxies is based on the refined SDSS $u-r$ vs. $r-z$ diagram \citep[e.g.,][]{Holden2012} designed for HSC galaxies as fully detailed in \citetalias{K21}. The usage of the $urz$ diagram instead of the traditional $UVJ$ diagram \citep{Williams2009} is because the $J$-band flux is not well constrained in our SED fitting as only optical photometry from HSC are input. As shown in the right panel, eFEDS AGNs show a high star-forming fraction at $\logm < 11.0\,\msun$ as also seen for SDSS quasars, while it drops at the high stellar mass end that resembles the inactive galaxy population. However, it is evident that eFEDS AGNs and SDSS quasars exhibit an excess in the star-forming fraction across a broad stellar mass range, except for low mass galaxies ($\logm<10.2\,\msun$) where all the three populations tend to be actively forming stars. 

We note that the cross-contamination between star-forming and quiescent classifications could be significant for both AGNs and inactive galaxies due to the limited photometric coverage as well as uncertainties and degeneracies in the SED fitting \citep{K21}. In particular, dusty star-forming galaxies could be misclassified as quiescent galaxies, especially at the high mass end \citep[e.g.,][]{Williams2009}. However, it is not likely that the inactive galaxy population contains a significantly higher dusty fraction that can erase the differences seen in Figure \ref{fig:uvm}, since AGN host galaxies are typically gas and dust rich \citep[e.g.,][]{Rosario2018, Shangguan2020, Yesuf2020}. Therefore, we conclude that AGNs show a preference for massive, star-forming host galaxies. This could be driven by the mutual dependence of BH growth and star formation on the cold gas reservoir. Our result is consistent with numerous studies that demonstrate the negative feedback from AGNs does not impose an instantaneous impact on the galaxy-wide star formation.

\subsection{Optical profile shapes (Sersic index distribution)}
\label{subsec:sersic_n}

\begin{figure*}
\includegraphics[width=\linewidth]{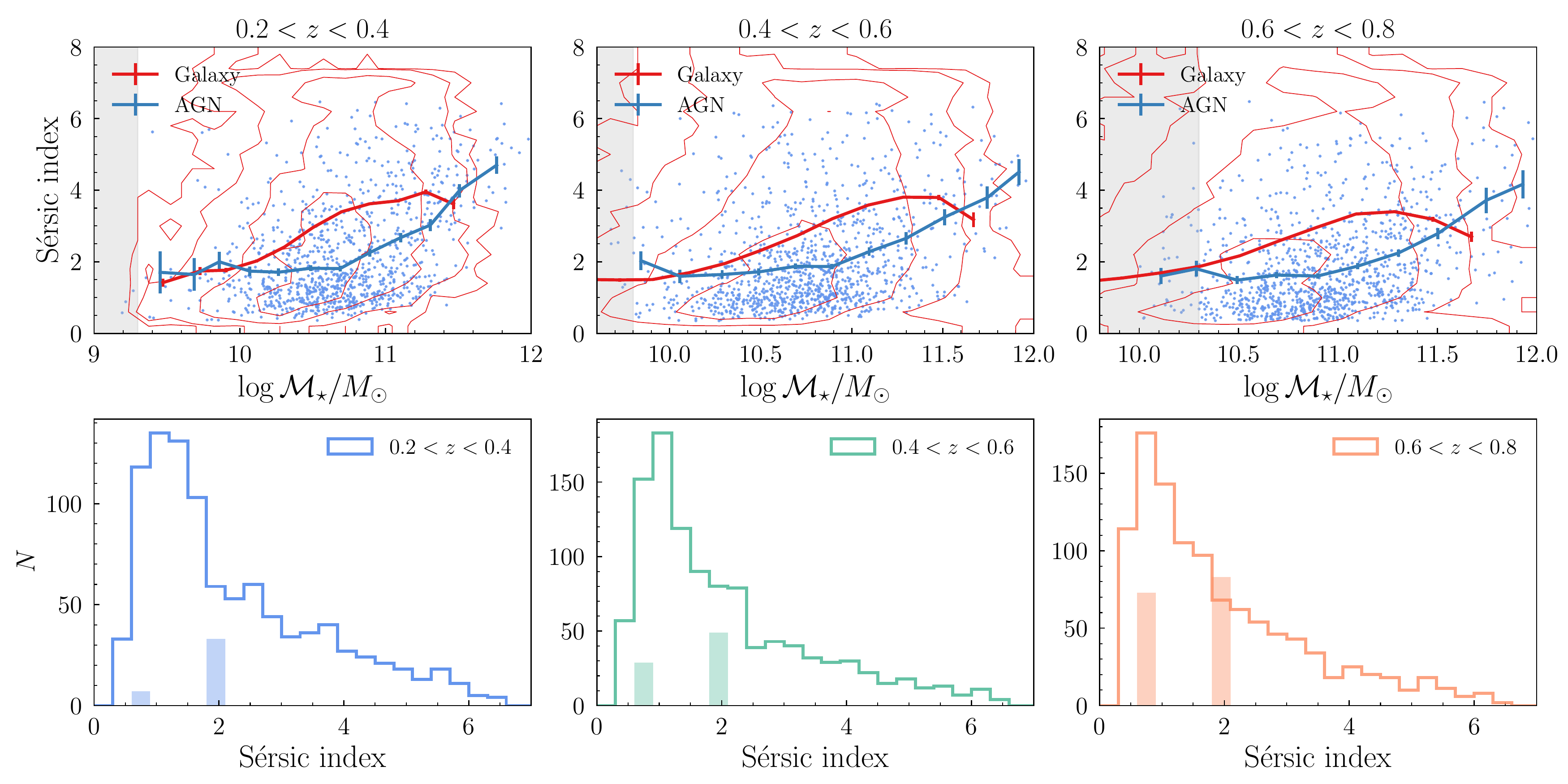}
\caption{Top: \sid~as a function of \m~for eFEDS AGNs (blue dots) and inactive galaxies (red contours) with each column presenting a different redshift interval as indicated. The average \ss~indices for the two populations are shown as red (galaxy) and blue (AGN) curves, respectively. The vertical shaded region marks our \m~cut.
Bottom: distribution of \sid~for eFEDS AGNs. Sources with fixed \ss~indices are shown as shaded bars at $n=0.7$ and $n=2.0$.}
\label{fig:sersic_n}
\end{figure*}

There has been much effort in the literature to determine which galaxy types AGNs prefer to reside in which can aid in our understanding of their means of growth and BH-galaxy coevolution.  However, previous studies have not yet achieved a coherent picture as both disk-dominated and bulge-dominated host morphology have been reported by different groups \citep[e.g.,][]{Schawinski2011, Cisternas2011a, Ding2020, L21, JLi2021, Zhuang2022}. To address this issue, we present the distribution of \sid~for our eFEDS AGN hosts to assess the prevalence of disk- ($n\sim1$) and bulge-dominated ($n\sim4$) systems.  The elevated host galaxy contribution to optical emissions enables us to measure the \ss~index of X-ray AGNs to higher accuracy which are challenge to constrain for luminous quasars with $\fr \lesssim 30\%$ or/and host-galaxy $i \lesssim 22$~mag, especially for systems with $n>4$ \citepalias{L21}. As shown in Figure \ref{fig:ratio}, 83\% of our eROSITA-HSC AGNs have $\fr>50\%$; their host galaxies are almost all brighter than 21 mag at $z<0.6$ and only a small fraction of them are fainter than 22~mag at $0.6<z<0.8$.  While constraining the \ss~index for individual sources is definitely more challenging than galaxy magnitude and size with HSC imaging \citep{Ishino2020, L21}, with such a bright host-galaxy, we can at least constrain the \ss~index for large samples in a statistical sense as verified in \citetalias{L21}. 

We note that recently, \cite{D22} compared the \ss~index measured from the HST ACS/WFC imaging and Subaru HSC imaging for AGNs in the COSMOS field and found strong inconsistency. This led them to conclude that the \ss~index measured from HSC imaging is meaningless even at $z<1$. However, although the superior spatial resolution and depth of HST allows \cite{D22} to measure host properties down to $\sim25$~mag in deep fields, those galaxies who are rarely brighter than $\sim20$~mag are not the typical host galaxies detected by wide area surveys such as combining HSC wide with SDSS and eROSITA, thus their result cannot be generalized to evaluate our sample. The fact that bright and massive galaxies are intrinsically larger (see Section \ref{subsec:size}) also makes our AGNs easier to decompose. Moreover, the wide stellar mass coverage and improved number statistics make it possible to split our sample into different \m~ranges and further quantify the subsample differences in the \ss~index distributions.

In Figure \ref{fig:sersic_n} we plot the \sid~vs. stellar mass relation for X-ray AGNs in the top panels and the \sid~distributions in the bottom panels, separated into three redshift intervals. For comparison, we also show in the top panels the inactive galaxies from \citetalias{K21}. Both measurements are based on the HSC $i$-band imaging which probes rest-frame wavelengths 5900~\AA, 5100~\AA, and 4500~\AA~in each interval, respectively. Overall, we find that the \ss~indices are consistent with AGNs residing in galaxies with significant stellar disks ($n<2$) at low to moderate stellar masses ($\logm < 11.0\,\msun$). As stellar mass approaching the ultramassive regime ($\logm > 11.0\,\msun$), AGN hosts become increasingly bulge dominated and quiescent (Section \ref{subsec:color}), although there may be a selection bias that AGNs hosted by massive star-forming disks could be obscured by their host galaxies thus undetected by eROSITA which mainly operates in the soft X-ray band (see Section \ref{subsec:nh}).

\begin{figure*}
\centering
\includegraphics[width=\linewidth]{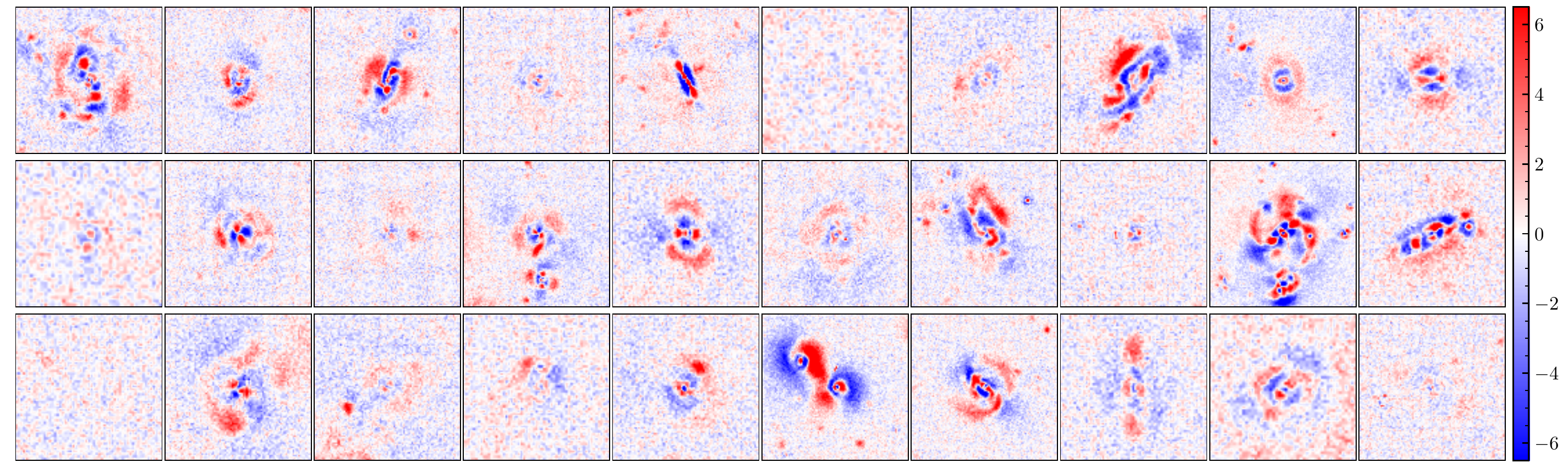}
\caption{Normalized $i$-band residual images (fitting residual divided by the error map) of 30 randomly-selected $z<0.4$ AGNs. The prevalence of disk structures is evident after subtracting the best-fit point source and smooth galaxy component.}
\label{fig:residual}
\end{figure*}

While the  \sid~vs. stellar mass relation of AGNs mimics the increasing trend found for inactive galaxies, the average \ss~indices of AGN hosts are lower by $\sim0.5-1.5$ than the general galaxy population over a broad stellar mass range, i.e., they are diskier\footnote{The average \ss~index of inactive HSC galaxies shows an unexpected decline at $\logm\gtrsim11.3$ which is likely due to photo-$z$ errors causing problematic SED fitting results. We have tested that comparing to inactive galaxies in the CANDELS field \citep{vanderwel2014} leads to consistent result although their \ss~indices are measured in the HST F125W band.}. This is not surprising given the elevated star-forming fraction of AGN host galaxies. Such a result suggests that disk-related non-axisymmetric structures (e.g., circumnuclear disk, spiral arm and ring), as commonly resolved in the residual images of our $z<0.4$ sample shown in Figure \ref{fig:residual} (see also \citealt{Nagele2022}), may play an important role in driving instabilities and transferring gas fuel to the nuclear region \citep[e.g.,][]{Bournaud2012}. 

\subsection{Host sizes}
\label{subsec:size}

\begin{figure*}
\includegraphics[width=\linewidth]{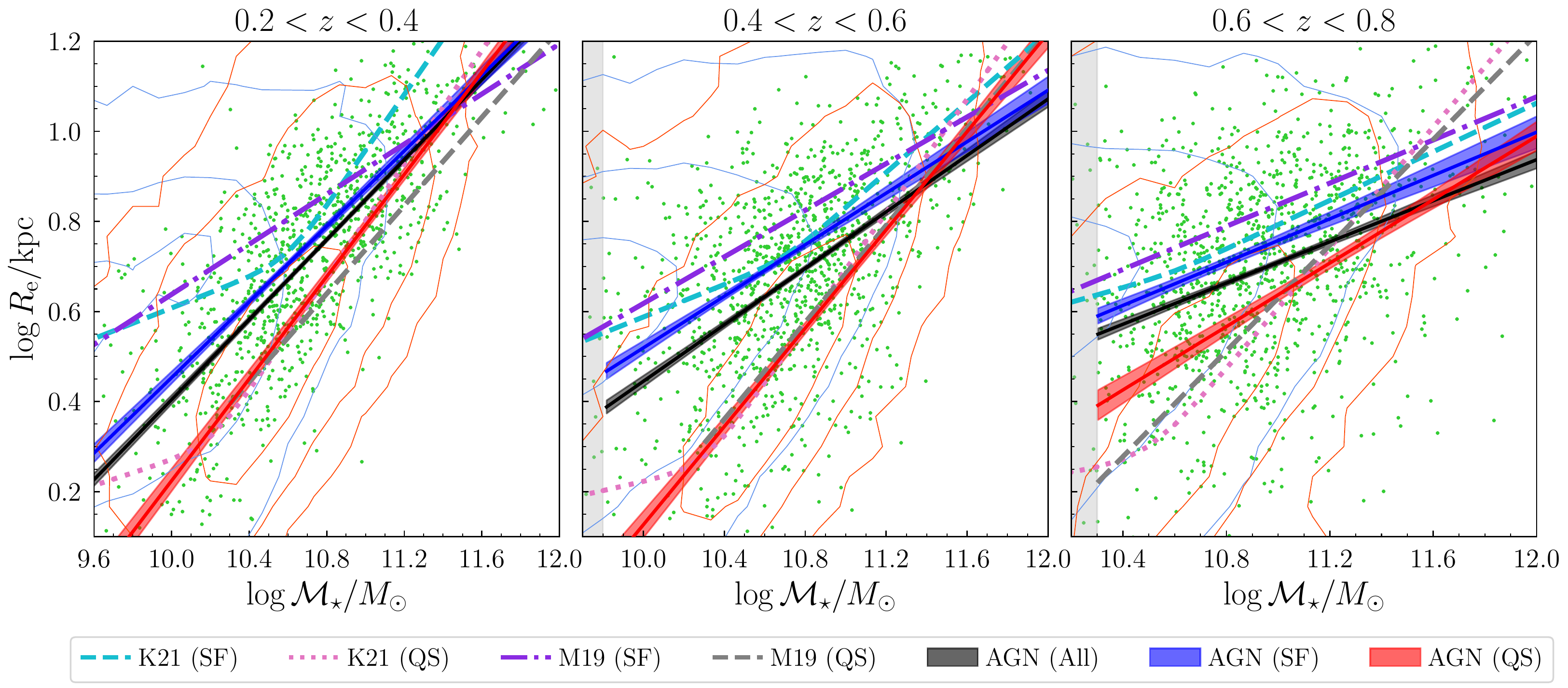}
\caption{Size-mass relations of AGN host galaxies compared to that of inactive galaxies in M19 and K21 (blue and red contours for inactive star-forming and quiescent galaxies, respectively) in three redshift intervals. The green points show measurements for individual AGNs. The best-fit size-mass relations in the form of single- or double- powerlaws of each populations are plotted and labeled. The vertical shaded region marks our \m~cut.}
\label{fig:mass_size}
\end{figure*}

Figure \ref{fig:mass_size} shows the size-stellar mass relation for AGN host galaxies. To compare the sizes of AGN hosts with the general galaxy population, we also plot the size-mass relations for inactive galaxies given by \cite{Mowla2019} (M19) and \citetalias{K21}. All sizes correspond to the semimajor axis of the \ss~model containing half of the light and have been corrected to a common rest-frame wavelength of 5000~\AA~following \cite{vanderwel2014}.
The size-mass relation given in \cite{Mowla2019} is fitted by a single-powerlaw model using galaxies in the CANDELS \citep{vanderwel2014} and COSMOS/DASH \citep{Mowla2019} fields, as shown by the purple dashed (for star-forming galaxies) and gray dash-dotted (for quiescent galaxies) lines. Note that their single-powerlaw fits do not capture the fast growth in size for massive star-forming galaxies at $\logm>11.0\,\msun$ (see their Figure 11). \citetalias{K21} find that the size-mass relation for HSC galaxies is better described by a double-powerlaw model. Their size-mass relations are plotted as cyan dashed (for star-forming galaxies) and pink dotted (for quiescent galaxies) curves, respectively.

Following \citetalias{L21}, we fit the size-mass relation of AGNs using a single-powerlaw model (with an intrinsic scatter $\sigma$) as we do not see a clear change in the powerlaw slope:
\begin{equation}
r(\m)/{\rm kpc} = A \times m_*^\alpha,
\label{eq:pwl}
\end{equation}
where $m_* \equiv \m/7\times10^{10}\,M_\odot$.
The best-fit parameters are summarized in Table \ref{table:fit}. In the fitting, we take the measurement uncertainties, the number density of low-mass and high-mass galaxies, and the cross-contamination between star-forming and quiescent hosts arised from the $urz$ classification into account through a maximum likelihood method \citep{vanderwel2014}. The cross contamination is corrected by assigning each AGN a weight based on the contamination fraction presented in \citetalias{K21}. Their contamination fraction is evaluated by taking the $UVJ$ classification of galaxies from the COSMOS/UltraVISTA, UKIDSS, UDS, and NMBS surveys as a ground truth, although the $UVJ$ selection of quiescent galaxies also suffers a level of contamination from dusty star-forming galaxies \citep{Williams2009}.
According to \citetalias{K21}, up to $40\%$ of $urz$-selected massive ($\logm > 11.2\,\msun$) star-forming galaxies could actually be quiescent systems which are more compact, thus failing to account for this misclassification will bias the size assessment for star-forming hosts, i.e., underestimate their average sizes, and vice versa for quiescent hosts. 

\begin{table*}
\centering
\renewcommand{\arraystretch}{1.4}
\caption{Best-fit parameters of single-powerlaw fits in the form of Equation \ref{eq:pwl} to the size--mass relation for all AGNs, star-forming AGNs and quiescent AGNs down to the stellar mass cut. Column (1) samples used to perform analytic fits, (2) $z_{\rm med}$ is the median redshift, (3) ${\rm log}\,(\mathcal{M}_{\rm \star, cut}/M_\odot)$ is the stellar mass cut, (4, 5, 6) $A$, $\alpha$ and $\sigma$ are the intercept, slope and intrinsic scatter of the \re--\m~relation. The uncertainties on parameters which are less than 0.01 are indicated as 0.}
\begin{tabular}{cccccc}
\hline
\hline
Sample & $z_{\rm med}$ & ${\rm log}\,(\mathcal{M}_{\rm \star,cut}/M_\odot)$ & $A$ & $\alpha$ & $\sigma$\\
(1) & (2) & (3) & (4) & (5) & (6)\\
\hline
 &  0.3 & 9.3  & $6.03_{-0.07}^{+0.07}$ & $0.45_{-0.01}^{+0.01}$ & $0.16_{-0.00}^{+0.00}$ \\
All  &  0.5 & 9.8 & $5.13_{-0.07}^{+0.07}$ & $0.32_{-0.01}^{+0.01}$ & $0.19_{-0.00}^{+0.00}$ \\
 &  0.7 & 10.3 & $4.72_{-0.06}^{+0.07}$ & $0.23_{-0.02}^{+0.02}$ & $0.19_{-0.00}^{+0.00}$ \\
 \hline
  &  0.3 & 9.3 & $6.42_{-0.13}^{+0.13}$ & $0.42_{-0.02}^{+0.02}$ & $0.15_{-0.01}^{+0.01}$ \\
Star-forming  &  0.5 & 9.8 & $5.77_{-0.12}^{+0.12}$ & $0.29_{-0.02}^{+0.02}$ & $0.16_{-0.01}^{+0.01}$ \\
 &  0.7 & 10.3 & $5.26_{-0.12}^{+0.12}$ & $0.24_{-0.03}^{+0.03}$ & $0.18_{-0.01}^{+0.01}$ \\
 \hline
  &  0.3 & 9.3 & $5.05_{-0.14}^{+0.15}$ & $0.57_{-0.02}^{+0.02}$ & $0.14_{-0.01}^{+0.01}$ \\
Quiescent  &  0.5 & 9.8 & $3.86_{-0.10}^{+0.11}$ & $0.55_{-0.03}^{+0.03}$ & $0.18_{-0.01}^{+0.01}$ \\
 &  0.7 & 10.3 & $3.82_{-0.14}^{+0.15}$ & $0.35_{-0.04}^{+0.03}$ & $0.20_{-0.01}^{+0.01}$ \\
\hline
\end{tabular}
\label{table:fit}
\end{table*}

Consistent with \citetalias{L21}, we see a clear size-mass relationship for AGN host galaxies as shown by the solid lines in Figure \ref{fig:mass_size}. The average sizes for AGN hosts tend to lie between those of inactive star-forming and quiescent galaxies over all redshifts considered. When isolating AGNs in star-forming and quiescent hosts, we see a clear size difference at $\logm<11.5\,\msun$ that quiescent hosts tend to be more compact, which resembles the size difference seen for inactive galaxies. This validates that the $urz$ classification can indeed distinguish star-forming and quiescent galaxies.
The sizes of star-forming hosts tend to be smaller than inactive star-forming galaxies at a given stellar mass, while still being larger than inactive quiescent galaxies in an average sense. Combining with previous sections, our result indicates that the bulk of X-ray AGNs with moderate stellar masses prefer to lie in compact star-forming disks, which agrees with recent findings in \citetalias{L21} and a study of 32 X-ray selected broad-line (type 1) AGNs at $1.2<z<1.7$ based on HST/WFC3 imaging in \cite{Silverman2019}. As fully described in \citetalias{L21}, this result suggests that the physical mechanism responsible for building the central mass concentration can also efficiently trigger BH accretion, as also seen in cosmological simulations \citep[e.g,][]{Habouzit2019}. This could be caused by a compaction event (driven by internal disk instabilities or external mergers) whereby gas can be transported to the nuclear region to both grow the BH and the bulge \citep[e.g.,][]{Dekel2014}. 

On the other hand, the size-mass relation of quiescent hosts closely follows that of inactive quiescent galaxies, except for the most massive hosts at $z>0.6$ where they lie significantly below. It is well established that the fast size growth of massive quiescent galaxies is driven by dry minor mergers \citep[][]{Naab2009, vanderwel2014, K21}. This may indicate that high-redshift quiescent AGN hosts have just formed their compact cores after a compaction-then-quenching event \citep[e.g.,][]{Zolotov2015}, but have not yet grow their sizes through subsequent dry mergers.

We end this section by emphasizing that measuring the intrinsic size-mass relation involves complex simulations and corrections to account for systematic measurement biases, contamination between star-forming and quiescent galaxies,  catastrophic photo-$z$ failures, and so on \citep[e.g.,][]{K21, L21}. The (sometimes subtle) size difference between AGNs and inactive galaxies could still be driven by the unknown systematics from ground-based imaging (e.g., systematic difference in the photometry and stellar mass for different surveys; \citealt{Tanaka2015, L21, K21}). In the upcoming decade, the synergy between eROSITA, Euclid, LSST, NGRST, and the Chinese Space Station Telescope will deliver unprecedented imaging data in the UV, optical and IR bands of billions of galaxies and millions of AGNs. Improved measurements based on these datasets and the state-of-the-art image modeling tool will push the current analysis to higher redshifts with a much larger sample and firmly put AGNs in the context of galaxy evolution.

\subsection{Host-galaxy contribution to AGN obscuration and sample biases}
\label{subsec:nh}

\begin{figure*}
\includegraphics[width=\linewidth]{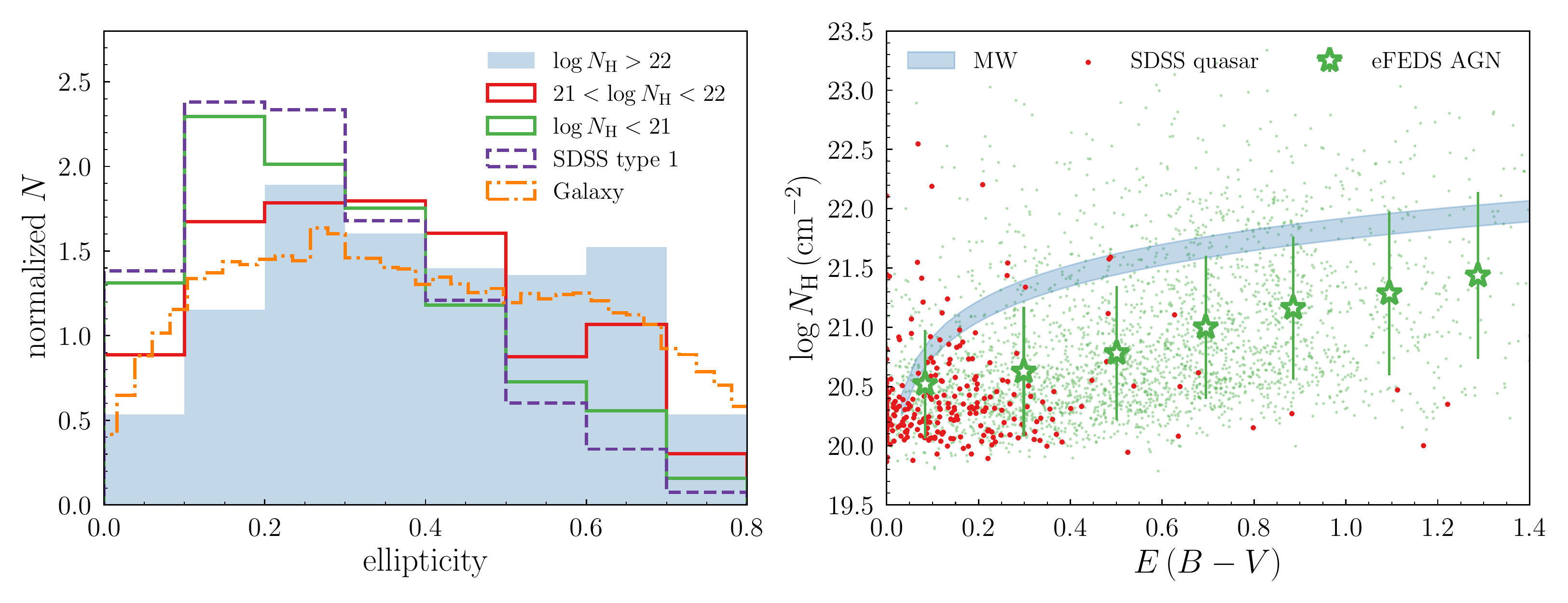}
\caption{Left: distributions of galaxy ellipticity for eFEDS AGNs in three \nh~intervals, SDSS quasars, and \m-matched inactive galaxies. Right: \nh~vs. $E(B-V)$ for eFEDS AGNs (small green points) with those being identified as type 1 quasars in the SDSS DR14 quasar catalog labelled by red points. Mean values and their dispersions are shown by open green asterisks. The blue shaded region gives the galactic dust-to-gas ratios from the literature ($\nh/\av=1.79-2.69\times10^{21}\rm\ cm^{-2}$ where $\av=3.1\times E(B-V)$).}
\label{fig:av_nh}
\end{figure*}

Motivated by past works on galaxy orientation and its contribution to nuclear extinction \citep[e.g.,][]{Keel1980, Maiolino1995, Gkini2021}, we show in Figure \ref{fig:av_nh} (left panel) the distribution of host-galaxy ellipticity, defined as $1-b/a$ where $a$ and $b$ are the major and minor axes of the \ss~model from our image modeling, for three \nh~intervals: $\nh<10^{21}\, \cm$ (X-ray unobscured, 2280 sources), $10^{21}<\nh<10^{22}$ (moderately obscured, 891 sources), and $\nh>10^{22}$ (obscured, 243 sources). SDSS quasars in \citetalias{L21} and \m-matched inactive galaxies in \citetalias{K21} are also plotted for comparison. Given that the redshift and stellar mass distribution for AGNs in each \nh~intervals are nearly identical, we assume their intrinsic shape in the 3D space are similar and interpret the axis ratio as an indicator of inclination angle, where $\epsilon=0$ and $\epsilon=1$ represent face-on and edge-on orientations, respectively\footnote{In principle, this definition is only valid for disk galaxies. We have tested that our result holds when limiting to disk-dominated ($n<2$) systems where the orientation can be better indicated by the axis ratio.}. Interestingly, the hosts of X-ray unobscured AGNs are shift towards face-on systems and follow the ellipticity distribution of SDSS quasars, consistent with their type 1/X-ray unobscured classification. On the contrary, the ellipticity distribution of X-ray obscured AGNs follows that of inactive galaxies and covers both face-on and edge-on systems. Moderately obscured AGNs have ellipticities lie between that of unobscured and obscured populations. The two-sample Kolmogorov-Smirnov test confirms that the ellipticity distribution of unobscured AGNs is significantly different from that of obscured AGNs and inactive galaxies with  $p$-values $\ll 0.001$. Our result is unlikely to be a bias that unobscured AGNs produce more contamination to the host-galaxy light which makes the galaxies appear round, since we are able to well recover the ellipticity in the high-$f_{\rm gal}$ regime (see Appendix in \citetalias{L21}). We also limit our experiment to the 2259 sources with reliable \nh~classifications, namely {\tt{NHclass=2}} for unobscured cases and {\tt{NHclass=4}} for well-measured \nh~and find consistent results.
The lack of hosts being close to edge-on for the unobscured/type~1 cases illustrates that the host galaxy can contribute to the X-ray obscuration up to (at least) $10^{22}\, \cm$ and also the optical extinction and reddening.

\begin{figure}
\includegraphics[width=\linewidth]{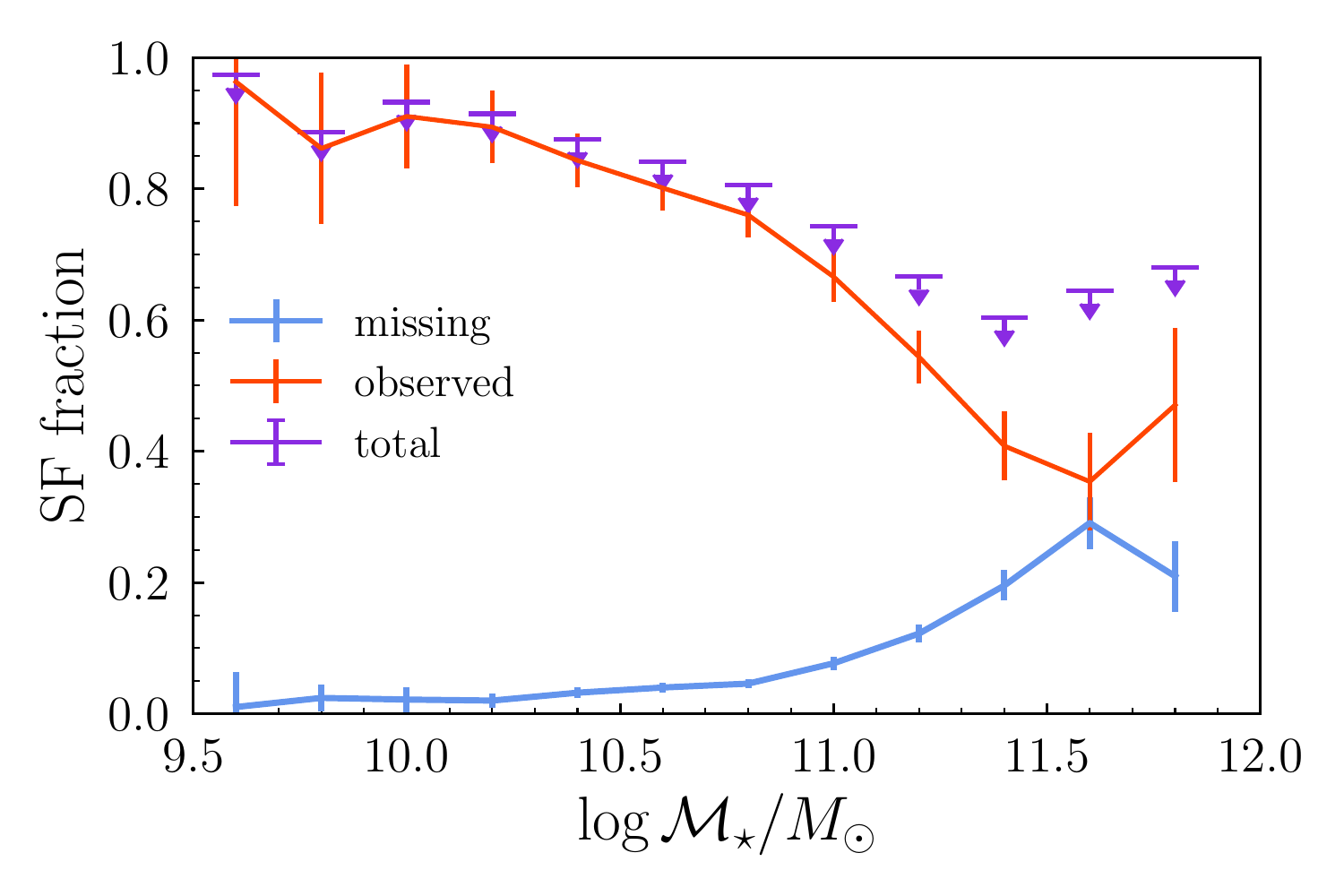}
\caption{{The missing (blue) and observed (red) star-forming host fraction as a function of stellar mass, averaged over $0.2<z<0.8$. The corrected total (observed+missing) star-forming fraction is shown in purple with the downward arrows indicating that the estimated missing fraction is an upper limit. The error bar is given by error propagation of Poisson uncertainty on $N_{\rm SF}$.}}
\label{fig:missing_sf_frac}
\end{figure}

To further investigate what causes the obscuration of the central point source, we derive the color excess $E(B-V)$ as a probe of optical extinction using our decomposed AGN SED in five HSC bands and explore the relationship between X-ray and optical obscuration. Specifically, we redden the standard type~1 quasar SED given in \cite{Lyu2017} by a range of $E(B-V)$ assuming the average Milkay Way extinction curve with $R_V=3.1$ \citep{Gordon2009}, and compare it with our observed AGN SED to derive the best-fit $E(B-V)$. Figure \ref{fig:av_nh} shows the comparison between \nh~and $E(B-V)$ for eFEDS AGNs. The galactic dust-to-gas ratio  \citep{Predehl1995, Nowak2012} is plotted as a blue shaded region. The eFEDS AGNs that are identified as type~1 quasars in the SDSS DR14 quasar catalog are highlighted in red. The exact value of $E(B-V)$ for each individual source depends on the assumed intrinsic AGN SED shape, thus we focus here on the average reddening (green asterisks).

We see a trend that the X-ray column density increases with optical extinction. The eFEDS AGNs have $E(B-V)$/\nh~close to or slightly higher than the galactic dust-to-gas ratio. Those identified as type 1 SDSS quasars are amongst the lowest \nh~and $E(B-V)$ in our sample, consistent with their optical classification of being unobscured along the line-of-sight. Similar result has been obtained for nearby type~1 AGNs \citep{Ogawa2021}. This is different from X-ray highly obscured AGNs ($\nh>10^{23}\,\cm$) which usually show a much lower $E(B-V)$/\nh, possibly due to an additional contribution of dust-free gas in the broad-line region to the heavy X-ray obscuration \citep[e.g.,][]{Maiolino2001, Burtscher2016, Ichikawa2019, Li2020}. 
Our result could further support that the obscuration in eFEDS AGNs with typically low \nh~may partly come from their host galaxies, with the dust properties similar to the galactic dust; or that a dusty outflow in the polar region could partially obscure the disk continuum while the compact X-ray emissions are leaked through the clumpy clouds without being absorbed, leading to higher values of $E(B-V)$/\nh~\citep[e.g.,][]{Asmus2019, Ogawa2021}.

If the host galaxy indeed contributes substantially to the X-ray obscuration, our sample may be biased against massive star-forming disks that are rich in gas and dust, as such AGNs would appear X-ray obscured thus missed by eROSITA. In fact, \cite{Buchner2017a} and \cite{Buchner2017b} examined the contribution of galactic-scale ISM (traced by the X-ray spectra of gamma ray burst) and AGN torus to the X-ray obscuration. They found that $\nh$ scales with $\m^{1/3}$ (see also \citealt{Lanzuisi2017}) and concluded that galactic-scale gas and dust is responsible for a large fraction of AGN obscuration in the Compton-thin regime. \cite{Goulding2012} further suggested that the host galaxy can produce $\nh$ as high as $10^{24}\,\cm$ by analysing the Si-absorption features in nearby Compton-thick AGNs with different galaxy inclination angles. Although due to the limited $\nh$ coverage we are unable to directly examine the $\nh-\m$ relation with our sample, the $\nh-\m$ relation in \cite{Buchner2017a} and \citealt{Lanzuisi2017} suggests that the average galaxy contribution to $\nh$ is about $\lognh \approx 21.5-22.0\,\cm$ at $\logm > 11.0\,\msun$. This is a factor of $\approx 3-10$ times higher than the average $\nh$ for our massive host galaxies ($\overline{\lognh}\approx21.0\,\cm$), although we caution that the uncertainties on \nh~measured from the eROSITA spectrum are large and only 588 sources in our sample have well-measured $\nh$ \citep{Liu2022}.

It is possible to estimate how many star-forming galaxies are missed by assuming that the intrinsic ellipticity distribution of AGN host galaxies follows that of inactive galaxies (in practice we assume $N_{\epsilon<0.4}/ N_{\epsilon>0.4}$ of AGNs equals to that of inactive galaxies) and attribute the deficiency in the high-ellipticity population to the undetected AGNs in the edge-on star-forming disks. The underlying assumption is that AGNs hosted by edge-on quiescent (gas- and dust-poor) galaxies should have been detected as unobscured AGNs by eROSITA provided that our line-of-sight is not intervened by the dusty torus. This gives us an upper limit on the missing star-forming host fraction\footnote{Here we are only estimating the missing star-forming fraction caused by the possible contribution of galactic disk to the nuclear obscuration instead of deriving an intrinsic star-forming fraction for the entire unobscured and obscured AGN (where the obscuration could come from the dusty torus) populations.}. As shown in Figure~\ref{fig:missing_sf_frac}, the missing star-forming fraction is negligible in the low mass regime but can be significant (up to 30\%) at the high mass end. The intrinsic star-forming fraction (averaged over $0.2<z<0.8$) can thus reach $\sim60\%$ even at $\logm > 11.4$. This source of bias against the detection of star-forming hosts must be bear in mind when interpreting the host galaxy properties of massive X-ray unobscured and optical type 1 AGNs.

\section{Discussion}
\label{sec:discussion}

\begin{figure*}
\centering
\includegraphics[width=0.9\linewidth]{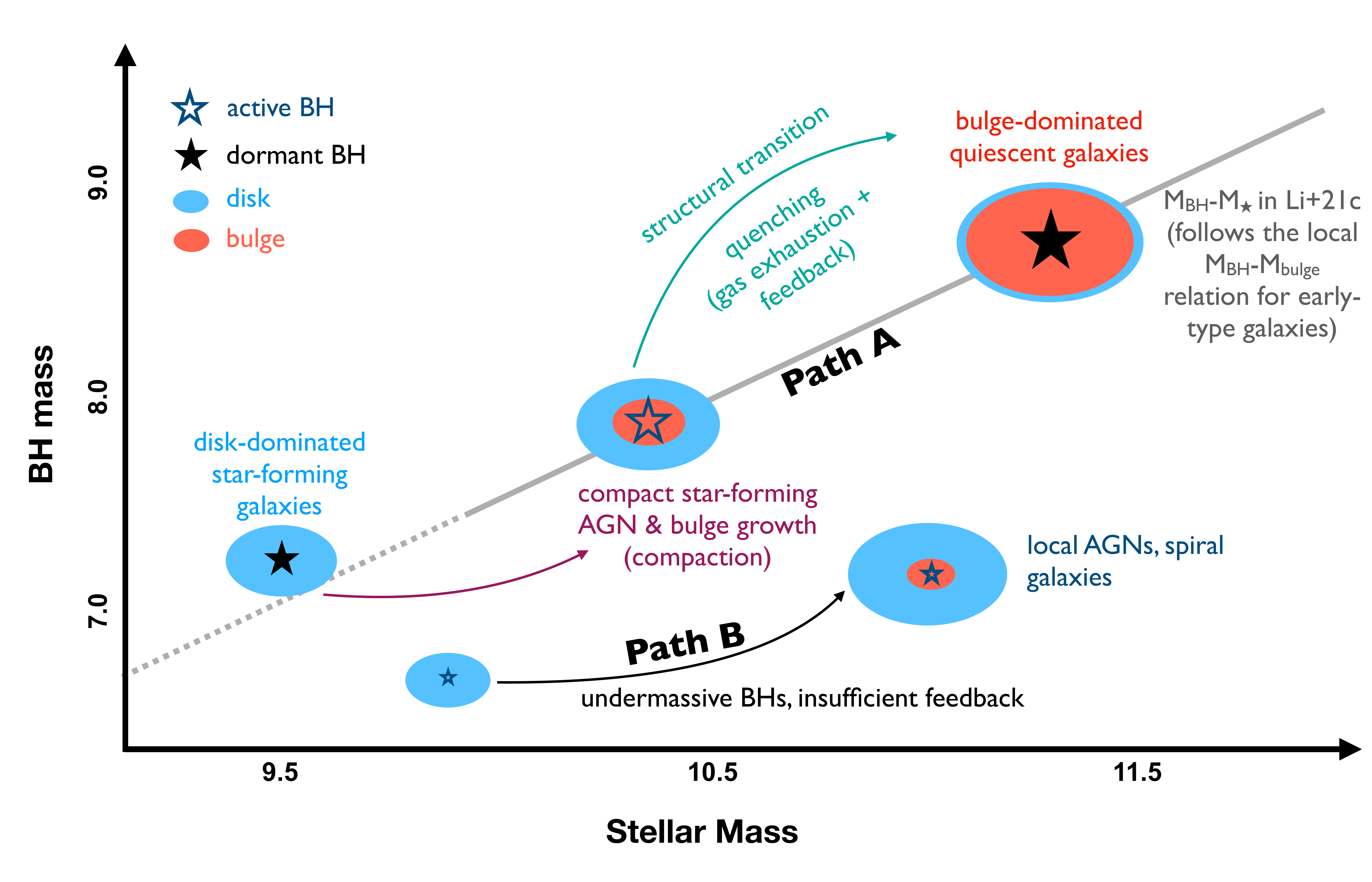}
\caption{A schematic view of the possible evolution pathways of SMBHs and their host galaxies that experience a structural transition and star-formation quenching through a compaction process (path A) or not (path B). See text in Section \ref{sec:discussion} for details.}
\label{fig:evolution}
\end{figure*}

The physical properties of AGN host galaxies provide key clues on the triggering mechanism of AGN activity and how BHs and galaxies coevolve. To this end, we have measured the structure, star formation and obscuration properties for a large sample of eROSITA-detected AGNs. We find that the host galaxies of low-\m~AGNs show a preference of disk-dominated star-forming galaxies and are on average more compact than inactive star-forming disks, while they become increasingly bulge-dominated and quiescent as \m~increases; although we caution that the star-forming disk fraction may be underestimated at the massive end due to the host galaxy contribution to nuclear obscuration.
The statistical difference between low-\m~($10.0 < \logm < 11.0$) and high-\m~($\logm > 11.0$) AGNs points to an evolution sequence in which the BH and galaxy grow in mass while transform in structure and star-forming activity, as desired to establish the BH -- bulge correlation seen in the local universe. In fact, \cite{Li2021b} found that the intrinsic (corrected for selection bias and measurement uncertainties) $\mbh-\m$ relation for massive systems ($\logm \gtrsim 10.0$ and $\logmbh \gtrsim 7.0$) does not evolve with redshift since $z\sim0.8$ and tightly follows the local $\mbh-\mbulge$ relation for early type galaxies. This led them to conclude that a structural transition event is required to transform the tight $\mbh-\m$ relation (probed by mostly disk-dominated broad-line AGNs) at high redshifts to the local $\mbh-\mbulge$ relation where $\mbulge\approx\m$ for massive galaxies (see also \citealt{Jahnke2009, Cisternas2011b, Schramm2013}). Such a result is further complemented by the analysis of the $\mbh-\sigma_\star$ (stellar velocity dispersion) relation in \cite{Silverman2022} where compact/quiescent AGNs are more closely aligned with the local relation. Given the bolometric luminosity of our sample (90\% have $43.5 < \loglbol < 45.5$) and their unobscured nature, their \mbh~are expected to be $\sim10^{6.5}-10^{8.5}\,\msun$ assuming an Eddington ratio of 0.1, thus they should generally follow the mass relation of broad-line AGNs. The correlations between \ss~index, galaxy size, and quiescent fraction with stellar mass found in this work provide supportive evidence of such an evolutionary sequence to establish the local scaling relations.

Even so, we notice that the majority of relatively low-\m~AGNs are still actively forming stars and have significant stellar disks. This is not necessarily in tension with the negative feedback scenario given the common dependence on the cold gas reservoirs for growing both BHs and galaxies. In fact, \cite{Ward2022} demonstrated that cosmological simulations (illustrisTNG, EAGLE, SIMBA) that incorporate strong BH feedback to quench massive galaxies also predict that AGNs with similar luminosities as this work tend to reside in gas-rich, star-forming galaxies. Nevertheless, for these AGNs to align with the scaling relations defined by early type galaxies, a transformation in their morphology and star-forming activity is still indispensable in the next few Gyrs. 

In the classical picture, major mergers and the subsequent AGN feedback are responsible to drive such a transformation \citep[e.g.,][]{DiMatteo2005, Hopkins2008}. However, although it remains controversial on whether AGNs show an excess in the merger rate compared to inactive galaxies, the overall major merger fractions among X-ray selected AGNs are low (a few percent up to $\sim30\%$; e.g., \citealt{Schawinski2011, Kocevski2015, Gao2020}). The prevalence of stellar disks in our sample is also at odds with the major merger scenario, but instead favors that internal disk instabilities or minor mergers are the dominant mechanisms in triggering AGN activity. Besides, most of the bulge growth seems to be not dominated by major mergers either, with simulations and observations demonstrating a greater role of violent disk instabilities in building classical bulges through a coalescence of massive disk clumps \citep[e.g.,][]{Elmegreen2008, Martig2012, Bell2017, Du2021}. This raises the question of whether the subsequent quenching (if any) is related with the AGN triggering and bulge formation process.

The lack of direct evidence of negative feedback in luminous AGNs can be reconciled with a significant time delay or if the cumulative energy injected by BHs in their low-accretion rate but long-lasting phase is more relevant in quenching star formation \citep[e.g.,][]{Weinberger2018, Piotrowska2022}. The latter scenario is supported by the observational evidence that the integrated BH mass appears to be the strongest predictor of galaxy quiescence in the local universe \citep[e.g.,][]{Terrazas2016, Piotrowska2022}. However, it is unclear how such a less violent feedback scenario is accompanied by a structural transformation given that quiescence is also well-correlated with galaxy compactness \citep[e.g.,][]{Whitaker2017}. It should also be realized that a prerequisite for a massive BH to quench a massive galaxy is that the system must have been gas rich and efficient in forming stars and fueling the BHs in order to accumulate such a large mass. One possible scenario is that the onset of quenching and structural transformation are both triggered by a dissipative compaction event which grows the bulge and rapidly consume the gas reservoirs in a few hundred million years through a central starburst \citep[e.g.,][]{Barro2016, Tacchella2016, Lapiner2021}. The feedback from BHs during or/and after the compaction event serves to further reduce the SFR and maintain the quiescence by preventing gas cooling and accretion. Given that the feeding of the central BHs is efficient during the compact star-forming phase \cite[e.g.,][]{Kocevski2017, Ni2021}, we can expect to see a high incidence of luminous AGNs in compact star-forming disks with concentrated molecular gas reservoirs \citep[e.g.,][]{Chang2020, Molina2021, Stacey2021}, a transformation of galaxy structure and star-forming activity along the growth of BHs and galaxies, and eventually a correlation between the integrated \mbh, the galaxy's central mass concentration and quiescence. A schematic view of this evolution pathway is presented in Figure \ref{fig:evolution} (Path A). 

While the above scenario can explain the statistical trends found in large AGN samples studied here and previous efforts based on HSC imaging \citep{L21, Li2021b, Nagele2022, Silverman2022}, it is crucial to emphasize that there could be significant variation in the evolution pathways of individual AGNs. Some galaxies may never undergo such transition events given the large dispersion in the size and \ss~index distributions. This is most likely to be the case for low luminosity AGNs ($\loglbol \lesssim 44.0$) which on average hosting low-mass BHs ($\logmbh \lesssim 7.0$) where the small \mbh~indicates that there may not have been much gas flowing into the nuclear region in the past. The structure transformation and feedback from BHs are likely inefficient in such systems and they will continue to grow the stellar disk (Path B). As a result, such AGNs will follow the mass relation of late type galaxies at $z=0$ that has a different normalization compared to early type galaxies \citep{Reines2015}. Alternatively, since the existing local dynamical $\mbh$ measurements are  biased to the most compact early-type galaxies \citep{Shankar2016}, the structure transformation of some AGNs, especially for high-\m~objects that are still hosted by pure disks at $z\sim0.2$, may not be necessary if massive BHs in disk galaxies exist in the local universe but their \mbh~are yet to be measured. A detailed discussion on how this bias may affect our evolutionary picture is beyond the scope of this paper.

\section{Conclusions}
\label{sec:summary}

In this work, we investigate the physical properties of X-ray selected AGNs and their host galaxies in the eFEDS field at $0.2<z<0.8$. Thanks to the wide area of eFEDS, a large sample of 3796 AGNs can be detected and analyzed. Using wide and deep optical imaging in the $grizy$ bands from the HSC-SSP survey, we decompose the AGN images into an underlying host galaxy (modeled by a 2D \ss~profile) and a point source component (modeled by a PSF), allowing us to measure the structural parameters for AGN host galaxies. SED fitting is performed on the decomposed host galaxy emission to estimate stellar population properties such as stellar mass and rest-frame colors. Combined with the X-ray spectral fitting result, we extensively investigated the properties of AGN host galaxies including star-forming activity, disk vs. bulge nature, galaxy size, and obscuration. Our main conclusions are the following:
\begin{enumerate}

\item The AGN contribution to the total optical light is broadly distributed at a given \lx~over all luminosities considered ($42.0<\loglx/\ergs<44.5$), and can be significant down to $\loglx\sim42.5\,\ergs$.  Ignoring the point source component can significantly bias the structural measurements of AGN host galaxies even at the lowest X-ray luminosities. This highlights the importance of AGN-host decomposition for the study of AGNs and their host galaxies in the optical bands (Section \ref{subsec:ratio}).

\item AGN host galaxies with low to moderate stellar masses ($\logm < 11.3\,\msun$) are predominately star-forming systems. The star-forming fraction drops at the high mass end while still showing an excess ($\sim20\%$) compared to inactive control galaxies (Section \ref{subsec:color}).

\item The \sid~distribution is consistent with the bulk of AGNs ($64\%$) residing in galaxies with significant stellar disks ($n<2$), while its relationship with stellar mass suggests that AGN hosts become increasingly bulge dominated and quiescent when approaching $\logm>11.0\,\msun$ (Section \ref{subsec:sersic_n}).

\item The size of AGN hosts, and for those classified as star-forming AGNs, tend to be more compact than inactive star-forming galaxies at a given stellar mass. This result is in agreement with studies of optically-selected SDSS quasars \citep{L21} and suggests that the physical mechanism responsible for building the central mass concentration can also efficiently trigger AGN activity (Section \ref{subsec:size}).

\item The host galaxies of X-ray unobscured ($\nh<10^{22}\,\cm$) AGNs are biased towards face-on systems, while X-ray obscured AGNs ($\nh>10^{22}\,\cm$) reside in both face-on and edge-on galaxies, indicating that the host galaxy can produce X-ray obscuration up to $10^{22}\, \cm$. The average $E(B-V)/\nh$ for the nuclei is similar to the galactic dust-to-gas ratio, lending further support that part of the obscuration could come from galaxy-scale gas and dust. This host galaxy contribution to the X-ray obscuration may cause a bias against the detection of massive star-forming disks as AGN host galaxies since such AGNs would appear X-ray obscured thus missed by eROSITA. The missing star-forming host fraction is negligible at $\logm<10.5\,\msun$ but can be significant (up to $30\%$) at the high mass end (Section \ref{subsec:nh}).

\end{enumerate}

\section*{Acknowledgements}
This work is based on data from {\it eROSITA}, the primary instrument aboard SRG, a joint Russian-German science mission supported by the Russian Space Agency (Roskosmos), in the interests of the Russian Academy of Sciences represented by its Space Research Institute (IKI), and the Deutsches Zentrum f\"ur Luft- und Raumfahrt (DLR). 
The SRG spacecraft was built by Lavochkin Association (NPOL) and its subcontractors, and is operated by NPOL with support from the Max-Planck Institute for Extraterrestrial Physics (MPE).
The development and construction of the {\it eROSITA} X-ray instrument was led by MPE, with contributions from the Dr. Karl Remeis Observatory Bamberg \& ECAP (FAU Erlangen-Nuernberg), the University of Hamburg Observatory, the Leibniz Institute for Astrophysics Potsdam (AIP), and the Institute for Astronomy and Astrophysics of the University of T\"ubingen, with the support of DLR and the Max Planck Society. 
The Argelander Institute for Astronomy of the University of Bonn and the Ludwig Maximilians Universit\"at Munich also participated in the science preparation for {\it eROSITA}. 
The {\it eROSITA} data shown here were processed using the eSASS/NRTA software system developed by the German {\it eROSITA} consortium. \\

The Hyper Suprime-Cam (HSC) collaboration includes the astronomical communities of Japan and Taiwan, and Princeton University.  The HSC instrumentation and software were developed by the National Astronomical Observatory of Japan (NAOJ), the Kavli Institute for the Physics and Mathematics of the Universe (Kavli IPMU), the University of Tokyo, the High Energy Accelerator Research Organization (KEK), the Academia Sinica Institute for Astronomy and Astrophysics in Taiwan (ASIAA), and Princeton University.  Funding was contributed by the FIRST program from the Japanese Cabinet Office, the Ministry of Education, Culture, Sports, Science and Technology (MEXT), the Japan Society for the Promotion of Science (JSPS), Japan Science and Technology Agency  (JST), the Toray Science  Foundation, NAOJ, Kavli IPMU, KEK, ASIAA, and Princeton University. JS is supported by JSPS KAKENHI Grant Number JP18H01251 and JP22H01262, and the World Premier International Research Center Initiative (WPI Initiative (WPI), MEXT, Japan. 
This paper makes use of software developed for the Large Synoptic Survey Telescope. We thank the LSST Project for making their code available as free software at  http://dm.lsst.org 
This paper is based [in part] on data collected at the Subaru Telescope and retrieved from the HSC data archive system, which is operated by Subaru Telescope and Astronomy Data Center (ADC) at NAOJ. Data analysis was in part carried out with the cooperation of Center for Computational Astrophysics (CfCA), NAOJ.\\

\section*{Data Availability}
The image decomposition and SED fitting results will be publicly available after the paper is accepted.



\bibliographystyle{mnras}
\bibliography{mnras_template.bbl} 


\bsp	
\label{lastpage}
\end{document}